\documentclass[a4paper,10pt,twocolumn]{article}
\usepackage[utf8]{inputenc}
\usepackage[T1]{fontenc}
\usepackage{xcolor,graphicx}
\usepackage{amsmath,amssymb,amsthm}
\usepackage{url}
\usepackage[final]{changes}
\definechangesauthor[name={First referee}, color={orange}]{R1}
\definechangesauthor[name={Second referee}, color={blue}]{R2}
\definechangesauthor[name={Authors}, color={gray}]{Authors}
\definechangesauthor[name={red}, color={red}]{red}
\setauthormarkup{}

\let\originalparagraph\paragraph
\renewcommand{\paragraph}[2][.]{\originalparagraph{#2#1}}
\providecommand{\keywords}[1]{\paragraph{Keywords} #1}
\usepackage{pbox}
\usepackage{float}
\newfloat{algo}{thp}{lop}
\floatname{algo}{Algorithm}

\usepackage{relsize} % define \smaller, used in \tilex
\definecolor{tileColor}{gray}{0.6}
\newcommand{\tilex}[1]{\colorbox{tileColor}{\color{white}{\text{\sf\smaller{#1}}}}}
\definecolor{lightgray}{gray}{0.95}

\newcommand\term[1]{\textbf{#1}\index{#1}}

\newcommand{\suchthat}                 {:}
                                       
\newcommand{\capacity}                 {C}
                                       
\newcommand{\pageSet}                  {\mathcal{P}}
\newcommand{\page}                     {p}
\newcommand{\pageIndex}                {k}
\newcommand{\pageIndexSet}             {P}
\newcommand{\pageCount}                {n}

\newcommand{\tileSet}                  {\mathcal{T}}
\newcommand{\tile}                     {t}
\newcommand{\tileIndex}                {j}
\newcommand{\tileIndexSet}             {T}
                                       
\newcommand{\symbSet}                  {\mathcal{A}}
\newcommand{\symb}                     {\alpha}
\newcommand{\symbIndex}                {i}
\newcommand{\symbIndexSet}             {A}
\newcommand{\symbA}                    {\alpha}
\newcommand{\symbB}                    {\beta}

\newcommand{\symbInTile}[2]            {a^{#1}_{#2}}
\newcommand{\symbInPage}[2]            {x^{#1}_{#2}}
\newcommand{\tileInPage}[2]            {y^{#1}_{#2}}
\newcommand{\pageUsed}[1]              {z_{#1}}
\newcommand{\fitness}                  {f}
\newcommand{\disparity}{d}

\newcommand{\multiplicity}[2]          {\mu_{#1}(#2)}
\newcommand{\card}[1]                  {\mathrm{Card}(#1)}
\newcommand{\volume}[1]                {\mathcal{V}(#1)}
\newcommand{\footprint}[2]          {|#2|_{#1}}
\newcommand{\queue}                    {\mathcal{Q}}

\newtheorem{proposition}{Proposition}

\newtheorem{simplifying_assumption}{Rule}
\newtheorem{conjecture}{Conjecture}
\newtheorem{corollary}{Corollary}

\theoremstyle{remark}
\newtheorem{definition}{Definition}
\newtheorem{remark}{Remark}
\newtheorem{example}{Example}
\begin{document}
	\title{Algorithms for the Bin Packing Problem\\ with Overlapping Items}
	\author{Aristide \textsc{Grange}$^*$ \and Imed \textsc{Kacem}$^*$ \and Sébastien \textsc{Martin}
		\thanks{\texttt{first\_name.surname@univ-lorraine.fr}, LCOMS EA7306, Université de Lorraine, Metz, FRANCE.}
	}
	\maketitle
	\begin{abstract}
We study
an extension of the bin packing problem,
where packing together two or more items
may make them occupy less volume than the sum of their individual sizes.
To achieve this property,
an item is defined as a finite set of symbols from a given alphabet.
Unlike the items of \textsc{Bin Packing},
two such sets can share zero, one or more symbols.
The problem was first introduced
in 2011 by Sindelar et al. \cite{Sindelar:2011}
under the name of \textsc{VM Packing}
with the addition of hierarchical sharing constraints
making it suitable for virtual machine colocation.
Without these constraints,
we prefer the more general name of \textsc{Pagination}.
After formulating it as an integer linear program,
we try to approximate its solutions with several families of algorithms:
from straightforward adaptations of classical \textsc{Bin Packing} heuristics,
to dedicated algorithms (greedy and non-greedy),
to standard and grouping genetic algorithms.
All of them are studied first theoretically,
then experimentally on an extensive random test set.
Based upon these data,
we propose
a predictive measure of the \added[id=R2]{statistical} difficulty of a given instance,
and finally recommend which algorithm should be used in which case,
depending on either time constraints or quality requirements.
\end{abstract}

	%!TEX root = /Users/aristide/Dropbox/pagination/paper/arXiv/pagination_CAIE.tex

\section{Introduction}

Over the last decade,
the \emph{book} (as a physical object) has  given ground
to the multiplication of screens of all sizes.
However,
the \emph{page} arguably remains the fundamental visual \replaced[id=R1]
    {unit}
    {unity}
for presenting data,
with degrees of dynamism varying
from video
to static images,
from infinite scrolling (e.g., Windows Mobile interface)
to semi-permanent display without energy consumption (e.g., electronic paper).
\textsc{Pagination} is to information
what \textsc{Bin Packing} is to matter.
Both ask how to distribute
a given set of items into the fewest number
of fixed-size containers.
But where \textsc{Bin Packing} generally handles concrete, distinct, one-piece objects,
\textsc{Pagination} processes abstract groups of data:
as soon as some data
are shared by two groups packed in the same container,
there is no need to repeat it twice.

\subsection{Practical applications}

As an introductory example,
consider the following problem.
A publisher offers
a collection of audio CDs for language learning.
Say that a typical CD consists of
100 short texts read by a native speaker;
for each of them,
a bilingual vocabulary of about 20 terms
has to be printed out on the CD booklet.
How best to do this?
The most expansive option,
both financially and environmentally,
would require the impression of a 100-page booklet,
i.e., with one page per audio text.
But now suppose that each page can accommodate up to fifty terms.
If all individual vocabularies are collated
into one single glossary,
no more than $100\times20/50=40$ pages are needed.
This is the cheapest option,
but the least convenient,
since it forces the consumer
to constantly leaf through the booklet
while listening to a given text.
To minimize cost without sacrificing usability,
the publisher will be better off to pack into each page
\replaced[id=R2]
    {as many}
    {the most}
individual vocabularies as possible.
If there were no common term between any two vocabularies,
this problem would be \textsc{Bin Packing};
but obviously,
most of the time,
the vocabulary of a given text
partially overlaps with several others:
this is what we propose to call \textbf{the pagination problem},
in short \textsc{Pagination}.
In this example,
it takes advantage of the fact that
the more terms are shared by two vocabularies,
the cheaper to pack them together;
as an added benefit,
a good pagination
will tend to group on the same page
the audio texts dealing with the same topic.

Coincidentally,
it was in this context of linguistics
that we first stumbled across \textsc{Pagination}.
At that time,
we needed to display
selected clusters of morphologically related
    \replaced[id=R1]
    {Chinese characters}
    {sinographs, or \emph{kanjis},}
on a pocket-sized screen.
A full description of our initial purpose
would be beyond the scope of this paper, 
and ultimately unnecessary,
since the problem is in fact perfectly general.
It only differs from \textsc{Bin Packing}
by the nature of the items involved:
instead of being atomic,
each such item is a combination of elements,
which themselves have two fundamental properties:
first,
they are all the same size
(relatively to the bin capacity);
and second,
their combination is precisely what
conveys the information we care about.

\begin{figure}[htbp]
    \includegraphics[width=\linewidth,trim={1cm 0.5cm 1cm 1cm}]{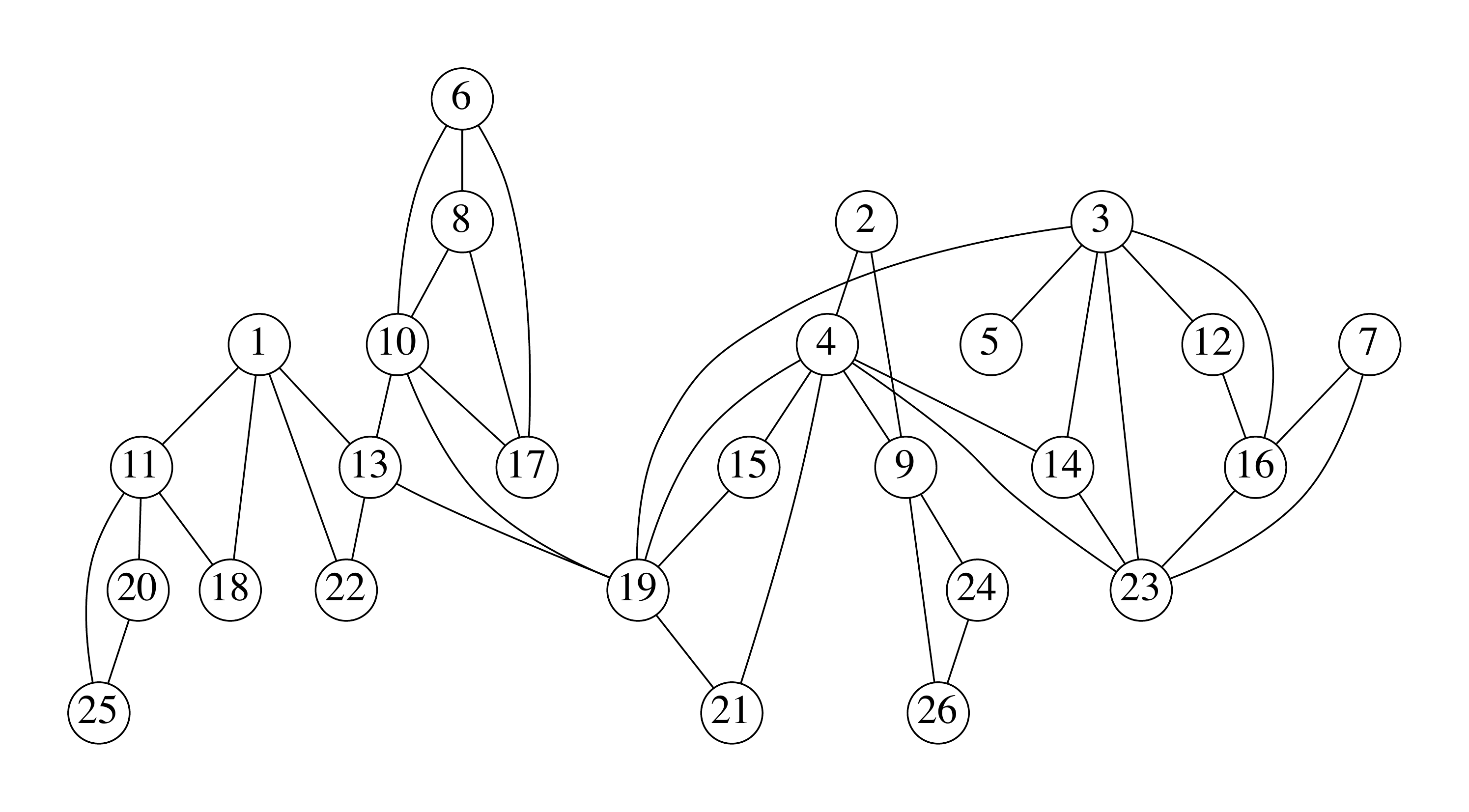}
    \includegraphics[width=\linewidth]{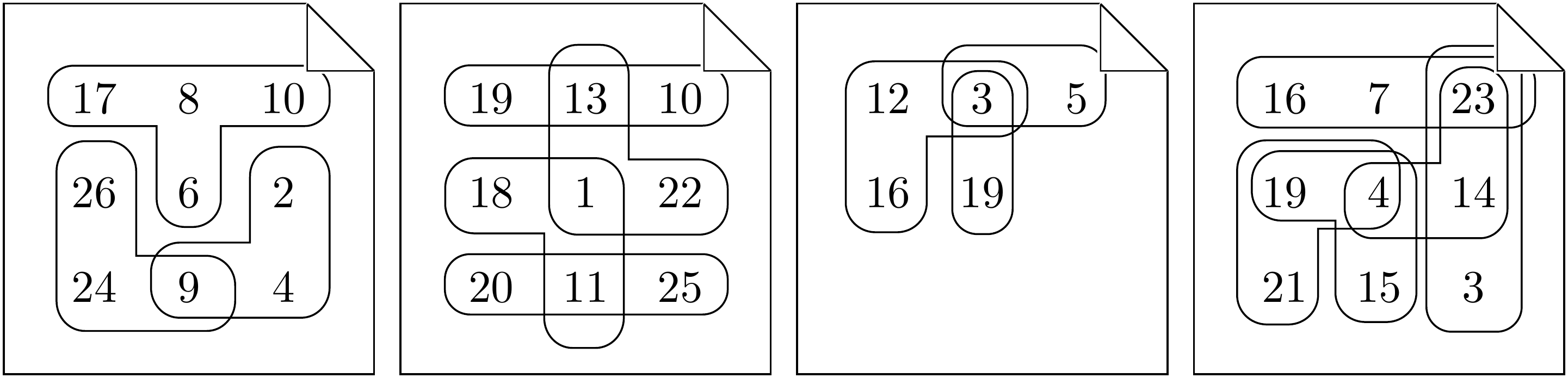}
    \caption{\label{fig:social_network}\it
    Application of \textsc{Pagination}
    to the visualization of the cliques of a given graph.
    The cliques are split over a small number of pages (here, 4)
    with a given capacity (here, at most 9 vertices).
    To allow each clique to be entirely contained in a single page,
    some vertices may be repeated on different pages
    (here, 19 on 3 pages, and 3, 4, 10, 16 on 2 pages).
    The individual page layouts are not relevant.
    }
\end{figure}

For instance,
the members of a social network
may interest us only to the extent that
they are part of one or several friendship circles.
Such groups of mutual friends are nothing more
than the so-called cliques of a graph
(Fig.~\ref{fig:social_network}, upper),
but the cliques are notoriously difficult to extract visually.
Visualizing them as separated sets of vertices is more effective,
although quite redundant.
A better compromise between compacity and clarity is attained by
paginating these sets as in Fig.~\ref{fig:social_network} (lower part).
Note that,
although no group is scattered across several pages,
finding all the friends of a given person
may require the consultation of several pages.

\subsection{Definition and complexity}

Our problem is not just about visualization.
It extends to any need of segmentation of partially redundant data
(see Section~\ref{sec:vm-allocation} for an application to virtual machine colocation).
Let us define it in the most general way:

\begin{definition}\label{def:pagination}
    \textsc{Pagination} can be expressed as the following decision problem:
\begin{itemize}
\item \emph{Input:}
    a finite collection $\tileSet$ of
    nonempty finite sets (the \term{tiles}\footnote{%
      The fact that \textsc{Pagination} generalizes \textsc{Bin Packing}
      has its counterpart in our terminology:
      the \emph{items} become the \emph{tiles},
      since they can overlap like the tiles on a roof.
    }) of \term{symbols},
    an integer $\capacity>0$ (the \term{capacity})
    and an integer $\pageCount>0$ (the \term{number of pages}).
\item \emph{Question:}
    does there exist an $\pageCount$-way partition
    (or \term{pagination})
    $\pageSet$
    of $\tileSet$ such that,
    for any tile set
    (or \term{page}\footnote{
      Likewise,
      the move from concrete to abstract
      is reflected by the choice of the term \emph{page} instead of \emph{bin}.
    })
    $\page$ of $\pageSet$,
    $|\cup_{\tile\in\page}\tile|\leq\capacity$?
\end{itemize}
\end{definition}

\begin{figure}[htbp]
\centering
\includegraphics[width=\linewidth]{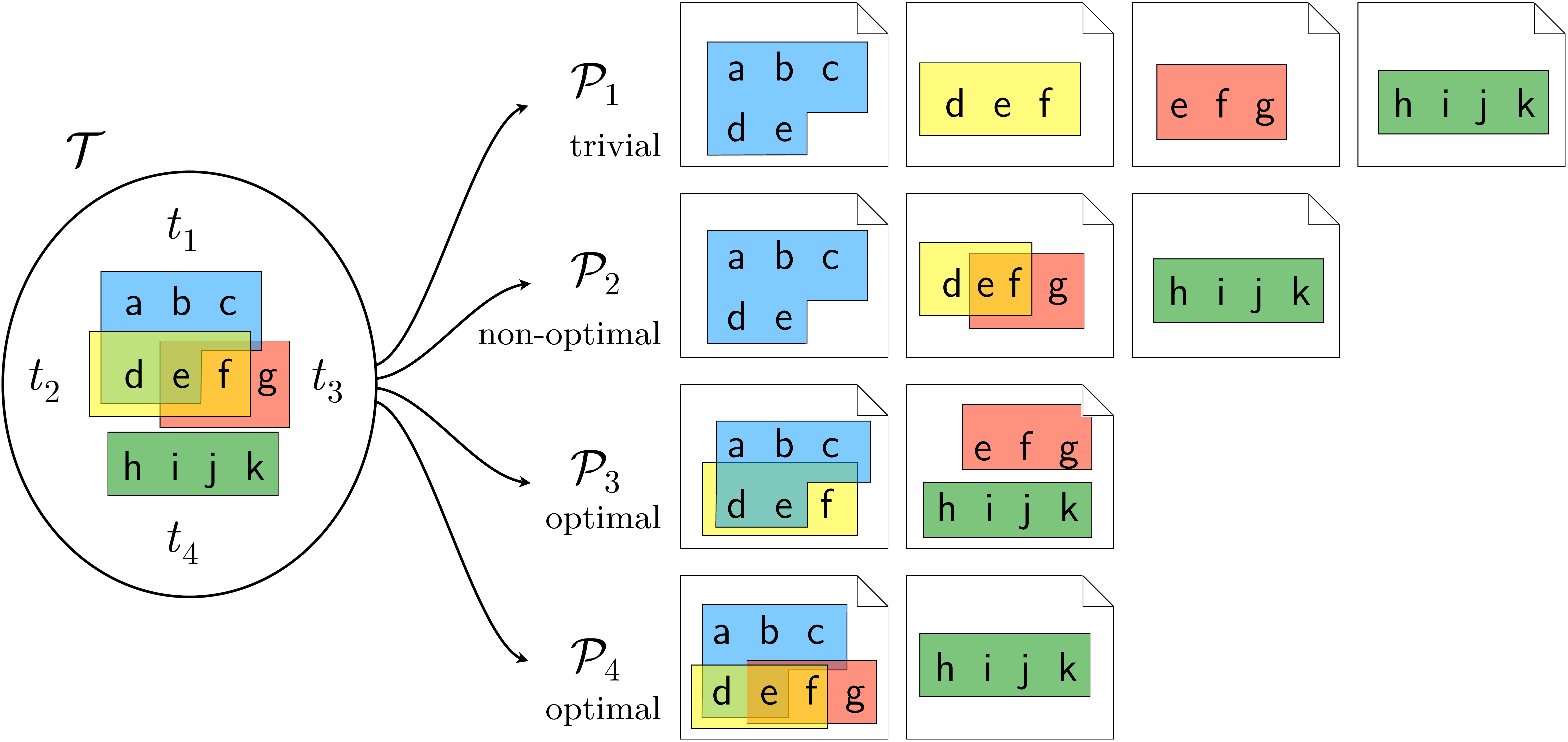}
\caption{\label{fig:intro}\it
For the tiles 
$\tileSet=\{
\tilex{abcde}\,\allowbreak
\tilex{def}\,\allowbreak
\tilex{efg}\,\allowbreak
\tilex{hijk}
\}$
and a capacity $\capacity=7$,
$\pageSet_3=(\{
\tilex{abcde}\,\allowbreak
\tilex{def}\},\allowbreak
\{\tilex{efg}\,\allowbreak
\tilex{hijk}\})$
is a possible optimal pagination
in $\pageCount=2$ pages.}
\end{figure}

\begin{example}
\replaced[id=R1]
    {On the left of Fig.~\ref{fig:intro},}
    {Figure~\ref{fig:intro} shows four valid paginations of the same set of tiles}
$\tileSet=\{\{\textsf{a},\textsf{b},\textsf{c},\textsf{d},\textsf{e}\}$,
$\{\textsf{d},\textsf{e},\textsf{f}\}$,
$\{\textsf{e},\textsf{f},\textsf{g}\}$,
$\{\textsf{h},\textsf{i},\textsf{j},\textsf{k}\}\}$
\added[id=R1]
    {denotes the tiles to be distributed on pages (with respect to a given capacity of 7).
    On the right, we present 4 possible such paginations of this set.}
For easier reading,
in the remainder of this paper,
any set of symbols (especially, a tile)
defined by extension
(for instance, $\{\textsf{w},\textsf{o},\textsf{r},\textsf{d}\}$)
will be represented as a gray block:
$\tilex{word}$.
Moreover,
in any collection
of symbol sets
(especially $\tileSet$, or a page),
the separating commas will be omitted:
$\{\tilex{May}$ $\tilex{June}$ $\tilex{July}\}$.
\end{example}

\begin{proposition}
\textsc{Pagination} is NP-complete in the strong sense.
\end{proposition}
\begin{proof}
Any given pagination $\pageSet$ can be verified in polynomial time.
Furthermore,
\textsc{Bin Packing}
is a special case of \textsc{Pagination}
with no shared symbol.
Hence,
the latter is at least as difficult as the former,
which is strongly NP-complete.
\end{proof}

In the rest of this paper,
we focus on the associated NP-hard optimization problem
(i.e., where the aim is to minimize the number of pages).

\subsection{Related works}

\added[id=R1]
{\textsc{Bin Packing} and its numerous variants
are among the most studied optimization problems,
and, as such, regularly subjected to comprehensive surveys
\cite{Coffman:1996, Coffman:2013, Johnson:1973, Martello:1990}.
The purpose of the present section is far more limited:
to extract from this vast literature the few problems
which exhibit the distinguishing feature of \textsc{Pagination},
i.e.,
which deal with objects able to overlap in a non-additive fashion.
For instance,
despite its similar name,
we will \emph{not} discuss \textsc{Newspaper Pagination},
a \textsc{Bin Packing} variant
studied in \cite{Lagus:1996}
for the placement of 
(obviously non overlapping)
blocks of text
on the columns of a generalized newspaper.}

\subsubsection{Virtual machine allocation}\label{sec:vm-allocation}
\textsc{Pagination} was first introduced in 2011 by Sindelar et al.\
\cite{Sindelar:2011}
in the context of virtual machine (VM) colocation.
A VM can be seen as a set of memory pages:
although most of them belong exclusively to one machine,
some pages are identical across several machines.
This happens more often as the configurations are close
in terms of platform, system version, software libraries, or installed applications.
If these VM run on the same physical server,
their common memory pages
can actually be pooled in order to spare resources.

The authors study two related problems.
The first one, \textsc{VM Maximization}, is defined as follows:
being given a collection of VM (our tiles),
each consisting in a set of memory pages (our symbols),
determine the most profitable subset of VM
that can be packed into a given set of servers (our pages)
having a capacity of $k$ memory pages (of $\capacity$ symbols).
Note that,
with one server,
this problem can be seen equivalently as
\textsc{Knapsack with Overlapping Items} \cite{Grange:2015},
or \textsc{Densest k-subhypergraph}.
Hajiaghayi et al.\ \cite{Hajiaghayi:2006}
have proven the latter to be hard
to approximate within a factor
of $2^{(\log n)^\delta}$
for some $\delta>0$,
and Rampersaud and Grosu \cite{Rampersaud:2014}
approximated it by a greedy heuristic
with an approximation ratio equal
to the number of VM.
Furthermore,
when the sharing among VM occurs in certain hierarchical fashions,
Sindelar et al.\ have used a left-right dynamic programming method
to devise a fully-polynomial time approximation scheme (FPTAS) \cite{Sindelar:2011}.

The second problem, \textsc{VM Packing},
is nothing else than \textsc{Pagination}:
it demands to allocate the VM
to a minimal number of servers.
Sindelar et al.\ have shown that
repeating their dynamic programming algorithm
solves it within a $O(\log n)$ factor
in a clustered version of the hierarchical model;
and that a greedy algorithm
achieves a factor of $3$
when each cluster is a singleton.

Although these hierarchical restrictions
lead to provably-good approximations algorithms,
they say nothing about the general model,
which the authors left open \cite{Sindelar:2011}.
Thereafter,
efforts have mainly focused
on models and methods designed specifically
for the purposes of solving real-world VM allocation problems
(e.g.,
\cite{Rampersaud:2016}
with multiple resource requirements in an online setting,
or \cite{Wilcox:2011}
for approaches based on genetic algorithms):
as far as we know,
the general case has been left unexplored.
In the interest of differentiating it
from these VM-oriented models,
we take the liberty to replace the terms of 
\textsc{VM packing}, memory pages, VM and servers,
by the application-agnostic terms of
\textsc{Pagination}, symbols, tiles and pages
(respectively).

\subsubsection{Hypergraph partitioning}\label{sec:hypergraph-partitioning}
Let us recall that a \term{hypergraph}
$G = (V, E)$
is defined by a set of vertices $V$
and a set of hyperedges $E$,
where each element of $E$ is a subset of $V$
\cite{Berge:1985}.
Bearing this in mind,
it is easy to see
the left-hand part of Fig.~\ref{fig:intro}
as a subset-based drawing
\cite{Kaufmann:2009}
of a hypergraph mapping the instance,
namely with
$V=\symbSet$ \replaced[id=R1]
    {(the set of vertices is the set of symbols)}
    {(the vertices are the symbols)}
and
$E=\tileSet$ (\replaced[id=R1]
    {(the set of hyperedges is the set of tiles).}
    {(the hyperedges are the tiles).}
Since \textsc{Pagination} is a partitioning problem,
it is natural to ask
whether we could take advantage
of the extensive literature on 
\textsc{Hypergraph Partitioning}.
In \cite{Karypis:1999} for instance,
the latter problem is defined as
“partitioning the vertices of a hypergraph into $k$ roughly equal parts,
such that a certain objective function defined over the hyperedges is optimized”.
Although
our capacity constraint on the symbols
is reminiscent of this “roughly equal” number of vertices,
the main purpose of \textsc{Pagination}
is to partition the tiles (i.e., the hyperedges),
and certainly not the symbols (i.e., the vertices).

So,
what happens if we try to exchange the roles
of the tiles and the symbols?
This gives the following alternative hypergraph representation
of the instances of \textsc{Pagination}:
$V=\tileSet$ (the vertices are the tiles)
and
$E=\symbSet$ (the hyperedges are the symbols).
To be more specific,
a symbol shared by several tiles connects them as a hyperedge,
while a proper symbol (i.e., a symbol belonging to one tile only)
connects this tile to itself as a hyperloop.
Now,
paginating $G$ indeed amounts to partitioning the vertices,
but in the meantime two issues have arisen.
First,
we do not care if each part contains roughly the same number of tiles:
we want instead that the number of involved hyperedges
is at most equal to $\capacity$.
Second,
we have to express our objective function (minimizing the number of pages)
on the hyperedges (the symbols).
To cite \cite{Karypis:1999} again:
“a commonly used objective function is to minimize the number of hyperedges that span different partitions”.
At first sight,
it would indeed seem reasonable to 
minimize the number of replications of a given symbol across the pages.
However,
this leads to another impasse:

\begin{proposition}\label{prop:min_replications_not_opt}
Minimizing the number of pages
and
minimizing the number of symbol replications
are not equivalent.
\end{proposition}
\begin{proof}[counterexample]
For
$\capacity=5$,
let
$\tileSet=
\{\tilex{a1}\,\allowbreak\tilex{357}\,\allowbreak\tilex{a2}\,\allowbreak\tilex{468}\}$.
The optimal pagination
$\{\{\tilex{a1}\,\allowbreak\tilex{357}\},\allowbreak\{\tilex{a2}\,\allowbreak\tilex{468}\}\}$
minimizes the number of pages (2),
but not the number of replicas (symbol \textsf{a} is replicated once).
Conversely,
the non-optimal pagination
$\{\{\tilex{a1}\,\allowbreak\tilex{a2}\},
\{\tilex{357}\},\allowbreak
\{\tilex{468}\}\}$
minimizes the number of symbol replications (0),
but not the number of pages (3).
\end{proof}

Therefore,
contrary to appearances,
\textsc{Pagination} has very little in common with \textsc{Hypergraph Partitioning}%
\footnote{
The problem studied in \cite{Duncan:1975},
and coincidentally named \emph{pagination}
(by reference to the fixed-length contiguous block of virtual memory,
or memory-pages),
is in fact a special case of \textsc{Hypergraph Partitioning},
where the objective is to minimize the total weight of edge-cuts in a weighted graph,
with an upper bound on the size of the parts.
}.

\bigskip

The rest of this paper is organized as follows.
In Section~\ref{sec:theoretical-tools},
we formulate \textsc{Pagination} as an integer linear program (ILP),
and introduce the various metrics and rules
used by our algorithms.
Then,
in Section~\ref{sec:heuristics},
we describe several heuristics and meta-heuristics for the problem.
Finally,
we compare the results produced by all the algorithms
(exact or not)
in Section~\ref{sec:experimental-results},
and conclude in Section~\ref{sec:conclusion}.

\section{Theoretical tools}\label{sec:theoretical-tools}

To look on our problem from another perspective,
let us formulate it as an ILP\added[id=R2]
    { problem}.
In addition,
we will be able
to solve some (admittedly simple) instances
with a generic optimization software.
Thereafter,
the introduction of several supplementary concepts
will permit us
to actually generate the instances,
and
to describe our own algorithms
for tackling the largest ones.

\subsection{Integer linear programming model}

\paragraph{Numberings}
We use the following sets of indexes:

\begin{itemize}
\item $\symbIndexSet=\{\symbIndex\suchthat\symbIndex=1,\dots,|\symbSet|\}$
        for the symbols;
\item $\tileIndexSet=\{\tileIndex\suchthat\tileIndex=1,\dots,|\tileSet|\}$
        for the tiles;
\item $\pageIndexSet=\{\pageIndex\suchthat\pageIndex\in\mathbb{N}\}$
        for the pages\footnote{
            For the sake of simplicity,
            we assume that
            an infinite number of pages are available.
            In practice,
            prior to the calculations,
            this number will be limited to a reasonable value,
            either given by a heuristic,
            or $|\tileSet|$ (one tile per page) in the worst case.
        }.
\end{itemize}

\paragraph{Inputs}
\begin{itemize}
\item $\capacity$ is \replaced[id=R2]
    {a strictly positive integer}
    {an integer nonnegative}
capacity;
\item $\symbInTile{\symbIndex}{\tileIndex}$ is an assignment of symbols to tiles:
$\forall\symbIndex\in\symbIndexSet,
\forall\tileIndex\in\tileIndexSet,
\symbInTile{\symbIndex}{\tileIndex}=1
$
if
$\symbIndex\in\tile_\tileIndex$,
and 0 otherwise.
\end{itemize}

\paragraph{Decision variables}
For any $\symbIndex\in\symbIndexSet$, $\tileIndex\in\tileIndexSet$ and $\pageIndex\in\pageIndexSet$,
we define:
\begin{itemize}
\item
$\symbInPage{\symbIndex}{\pageIndex}$
as equal to 1 if symbol $\symbIndex$ is present on page $\pageIndex$, and 0 otherwise
(pagination of the symbols);
\item
$\tileInPage{\tileIndex}{\pageIndex}$
as equal to 1 if tile $\tileIndex$ is present on page $\pageIndex$, and 0 otherwise
(pagination of the tiles);
\item
$\pageUsed{\pageIndex}$
as equal to 1 if page $\pageIndex$ is used, and 0 otherwise
(unitary usage of the pages).
\end{itemize}

It is worth noting that
$\symbInPage{\symbIndex}{\pageIndex} = \max_{\tileIndex\in\tileIndexSet}(\symbInTile{\symbIndex}{\tileIndex}\tileInPage{\tileIndex}{\pageIndex})$
and
$\pageUsed{\pageIndex} = \max_{\tileIndex\in\tileIndexSet}(\tileInPage{\tileIndex}{\pageIndex})$.
The mathematical model of \textsc{Pagination}
is thus entirely specified with $\tileInPage{\tileIndex}{\pageIndex}$:
the introduction of these auxiliary variables is only used
to achieve the linearity of its formulation.

\paragraph{Integer linear program}
A possible ILP formulation of \textsc{Pagination} is:
\begin{align}
\text{min.} 
& \sum_{\pageIndex\in\pageIndexSet} \pageUsed\pageIndex\notag \\
\text{s. t.}
\label{constraint:model:each_tile_is_placed}
& \sum_{\pageIndex\in\pageIndexSet} \tileInPage{\tileIndex}{\pageIndex}=1,
    & \forall\tileIndex\in\tileIndexSet
    \\
\label{constraint:model:used_pages_contain_symbols}
& \pageUsed{\pageIndex}\ge\symbInPage{\symbIndex}{\pageIndex},
    & \forall\symbIndex\in\symbIndexSet,
      \forall\pageIndex\in\pageIndexSet
    \\
\label{constraint:model:at_most_capacity_symbols_par_page}
& \sum_{\symbIndex\in\symbIndexSet} \symbInPage{\symbIndex}{\pageIndex} \le \pageUsed{\pageIndex}\capacity,
    & \forall\pageIndex\in\pageIndexSet
    \\
\label{constraint:model:tile_symbols_on_same_page}
& \symbInPage{\symbIndex}{\pageIndex} \ge \symbInTile{\symbIndex}{\tileIndex}\tileInPage{\tileIndex}{\pageIndex},
    & \forall\symbIndex\in\symbIndexSet,
      \forall\tileIndex\in\tileIndexSet,
      \forall\pageIndex\in\pageIndexSet
    \\
\label{constraint:model:type}
&
    \hspace{-1.65em} % fine-tuning for two columns
    \symbInPage{\symbIndex}{\pageIndex},\;\tileInPage{\tileIndex}{\pageIndex}, \pageUsed{\pageIndex}\;\text{all binary,}
    \hspace{-0.2em} % fine-tuning for two columns
    &
    \forall\symbIndex\in\symbIndexSet,
      \forall\tileIndex\in\tileIndexSet,
      \forall\pageIndex\in\pageIndexSet
\end{align}

Eq.~\ref{constraint:model:each_tile_is_placed} assigns each tile to exactly one page.
Eq.~\ref{constraint:model:used_pages_contain_symbols} ensures that a page is used as soon as it contains one symbol.
Conversely, thanks to the objective function,
every used page contains at least one symbol.
From Eq.~\ref{constraint:model:at_most_capacity_symbols_par_page},
a page cannot contain more than $\capacity$ symbols.
Eq.~\ref{constraint:model:tile_symbols_on_same_page} guarantees that,
when a tile belongs to a page, this page includes all its symbols%
\footnote{
    \replaced[id=R2]
    {Although Eq.~\ref{constraint:model:tile_symbols_on_same_page}
    produces a lot of unnecessary constraints of the form
    $\symbInPage{\symbIndex}{\pageIndex}\ge0$,
    we prefer it to the equivalent, but slightly less explicit formulation}
    {A tighter, but slightly less explicit formulation is still possible:}
    $
    \symbInPage{\symbIndex}{\pageIndex}\ge\tileInPage{\tileIndex}{\pageIndex},
    \forall\tileIndex\in\tileIndexSet,
    \forall\symbIndex\in\tile_\tileIndex,
    \forall\pageIndex\in\pageIndexSet
    $.
}.
The integrality constraints of the auxiliary variables
of Eq.~\ref{constraint:model:type}
could be relaxed.

\subsection{Counting of the symbols}

\begin{definition}[metrics]\label{def:metrics}
    Let
    $\symb$ be a symbol,
    $\tile$ a tile
    and
    $\page$ a set of tiles (most often, a page). Then:
    \begin{enumerate}
    \item
        the \term{size} $|\tile|$ of $\tile$
        is its number of symbols;
    \item
        the \term{volume} $\volume{\page}$ of $\page$
        is its number of distinct symbols:
        $\volume{\page}=|\cup_{\tile\in\page}\tile|$;
        and its complement
        $\capacity-\volume{\page}$,
        the \term{loss} on $\page$;
    \item
        by contrast,
        the \term{cardinality} $\card{\page}$
        is the total number of symbols (distinct or not) in $\page$:
        $\card{\page}=\sum_{\tile\in\page}|\tile|$.
    \item
        the \term{multiplicity} $\multiplicity{\page}{\symb}$
        counts the occurrences of $\symb$ in the tiles of $\page$:
        $\multiplicity{\page}{\symb}=
        |\{\tile\in\page\suchthat\symb\in\tile\}|$;
    \item
        the \term{relative size} of $\tile$ on $\page$
        is the sum of the reciprocals of the multiplicities of the symbols of $\tile$ in $\page$:
        $\footprint{\page}{\tile} =
        \sum_{\symb\in\tile}\frac{1}{\multiplicity{\page}{\symb}}$.
    \end{enumerate}
\end{definition}

In recognition of the fact
that a given symbol may occur several times on the same page,
the terms \emph{cardinality} and \emph{multiplicity} are borrowed
from the multiset theory \cite{Syropoulos:2001}.

\begin{example}
    In pagination $\pageSet_3$ of Fig.~\ref{fig:intro}:
size $|\tile_1|=5$,
volume $\volume{\page_1}=6$ (loss: $7-6$),
cardinality $\card{\page_1}=5+3$,
multiplicity $\multiplicity{\page_1}{\textsf{e}}=2$,
relative size $\footprint{\page_1}{\tile_2}=1/2+1/2+1$.
\end{example}

These definitions \replaced[id=R2]
    {are extended to several pages (\emph{i.e.}, sets of sets of tiles)}
    {can be extended to more than one page}
by summing the involved values:

\begin{example}
In the same figure,
volume $\volume{\pageSet_3}=6+7$ (loss: $7-1+7-7$),
cardinality $\card{\pageSet_3}=\card{\tileSet}=5+3+3+4$,
multiplicity $\multiplicity{\pageSet_3}{\textsf{e}}=\multiplicity{\tileSet}{\textsf{e}}=2+1$,
relative size $\footprint{\pageSet_3}{\tile_2}=\footprint{\tileSet}{\tile_2}=1/2+1/3+1/2$.
\end{example}

We are now able to express
a first interesting difference with \textsc{Bin Packing}:

\begin{proposition}\label{prop:min_loss_not_opt}
A pagination whose loss is minimal is not necessarily optimal.
\end{proposition}
\begin{proof}[counterexample]
For $\capacity=4$, the tile set $\tileSet=
\{\tilex{12}\,\allowbreak\tilex{13}\,\allowbreak\tilex{23}\,\allowbreak\tilex{ab}\,\allowbreak\tilex{ac}\,\allowbreak\tilex{bc}\}$
has a loss of $1+1$ on the optimal pagination
$(\{\tilex{12}\,\allowbreak\tilex{13}\,\allowbreak\tilex{23}\},\allowbreak \{\tilex{ab}\,\allowbreak\tilex{ac}\,\allowbreak\tilex{bc}\})$;
but a loss of $0+0+0$ on the non-optimal pagination
$(\{\tilex{12}\,\allowbreak\tilex{ab}\},\allowbreak \{\tilex{13}\,\allowbreak\tilex{ac}\},\allowbreak\{\tilex{23}\,\allowbreak\tilex{bc}\})$.
\end{proof}

As seen in this counterexample,
for the complete page set,
multiplicity, cardinality and relative size
do not depend on the pagination.
Then:

\begin{proposition}\label{prop:opt_eq_max_avg_weight}
A pagination is optimal if and only if the average cardinality of its pages is maximal.
\end{proposition}
\begin{proof}
Since the sum of the page cardinalities
is always equal to $\sum_{\tile\in\tileSet}|\tile|$,
its average depends only of the size $\pageCount$ of the pagination.
Hence, minimizing this size or that average is equivalent.
\end{proof}

\subsection{Simplifying assumptions}\label{sec:rules}

Definition~\ref{def:pagination}
encompasses many instances
whose pagination is either
infeasible (e.g., one tile exceeds the capacity), 
trivial (e.g., all tiles can fit in one page)
or reducible (e.g., one tile is a subset of another one).
The purpose of this subsection 
is to bring us closer to the core of the problem,
by ruling out
as many such degenerated cases as possible.
For each one,
we prove that
there is nothing lost
for an offline algorithm
to ignore the corresponding instances.

\begin{simplifying_assumption}\label{rule:no_inclusion}
No tile is included in another one:
% $\forall(\tile,\tile')\in\tileSet^2,\tile\subseteq\tile'\Leftrightarrow\tile=\tile'$.
$\forall(\tile,\tile')\in\tileSet^2\suchthat \tile\neq\tile',\tile\not\subseteq\tile'$.
\end{simplifying_assumption}

\begin{proof}
If $\tile\subseteq\tile'$, then $\tile$ and $\tile'$ can be put together on the same page of the optimal solution.
\end{proof}

\begin{remark}
    This does not hold for an online algorithm.
    Take for instance
    $\tileSet=\{\tilex{12}\,\allowbreak\tilex{345}\,\allowbreak\tilex{126}\,\allowbreak\tilex{378}\}$
    and $\capacity=5$.
    If the tiles are presented in this order,
    the first two may be placed on the first page,
    making necessary to create a new page for each remaining tile:
    $\{\{\tilex{12}\,\allowbreak\tilex{345}\}, \{\tilex{126}\}, \{\tilex{378}\}\}$.
    An optimal pagination would require two pages only:
    $\{\{\tilex{12}\,\allowbreak\tilex{126}\}, \{\tilex{346}\,\allowbreak\tilex{378}\}\}$.
\end{remark}

In other words,
the tile sets we deal with are Sperner families \cite{Sperner:1928}. It follows that:

\begin{corollary}[Sperner's Theorem]
	A page of capacity $\capacity$ contains at most $\binom{\capacity}{\lfloor\capacity/2\rfloor}$ tiles.
\end{corollary}

\begin{simplifying_assumption}\label{rule:no_total_cover}
	No tile contains all the symbols: $\nexists\tile\in\tileSet\suchthat\tile=\symbSet$.
\end{simplifying_assumption}
\begin{proof} Direct consequence of Rule~\ref{rule:no_inclusion}.
\end{proof}

\begin{simplifying_assumption}\label{rule:tile_size_upper_bound}
Each tile has less than $\capacity$ symbols: $\forall \tile\in\tileSet, |\tile|<\capacity$.
\end{simplifying_assumption}
\begin{proof}
    Let $\tile$ be an arbitrary tile.
    If $|\tile|>\capacity$, the problem has no solution.
    If $|\tile|=\capacity$,
    then no other tile $\tile'$ could appear on the same page as $\tile$
    without violating Rule~\ref{rule:no_inclusion}.
    Let $\pageSet'$ be an optimal pagination
    of the reduced instance $\tileSet\setminus\{\tile\}$.
    Then $\pageSet=\pageSet'\cup\{\{\tile\}\}$
    is an optimal pagination of $\tileSet$.
\end{proof}

\begin{simplifying_assumption}\label{rule:no_common_symbol}
No symbol is shared by all tiles:
$\nexists\symb\in\symbSet\suchthat\forall\tile\in\tileSet,\symb\in\tile$.
\end{simplifying_assumption}
\begin{proof}
    Otherwise,
    let $\pageSet'$ be an optimal pagination,
    for a capacity of $\capacity-1$,
    of the reduced instance $\tileSet'=\{\tile\setminus\symb\suchthat\tile\in\tileSet
    \text{~and~}t\neq\{\alpha\}\}$.
    Then adding $\symb$ to each page of $\pageSet'$
    gives an optimal pagination $\pageSet$ of $\tileSet$ for capacity $\capacity$.
    Should a tile $\{\alpha\}$ exist in $\tileSet$,
    it can be put back at no cost on any page of $\pageSet$.
\end{proof}

In other words, $\tileSet$ has not the Helly property \cite{Dourado:2009}.
Contrast this with \textsc{VM Packing} \cite{Sindelar:2011},
where the mere existence of \emph{root} symbols violates this rule.

\begin{simplifying_assumption}\label{rule:no_useless_symbol}
Each symbol belongs to at least one tile:
$\forall\symb\in\symbSet,\exists\tile\in\tileSet\suchthat\symb\in\tile$.
\end{simplifying_assumption}
\begin{proof}
    By Definition~\ref{def:pagination} of an instance.
\end{proof}

\begin{simplifying_assumption}\label{rule:coexistence}
Each tile is compatible with at least another one:
$\forall\tile\in\tileSet,
\exists\tile'\in\tileSet\setminus\{t\}
\suchthat |\tile\cup\tile'|\le\capacity$.
\end{simplifying_assumption}
\begin{proof}
    If there exists a tile $\tile$ not compatible with any other,
    any solution should devote a complete page to $\tile$.
    The conclusion of 
    the proof of Rule~\ref{rule:tile_size_upper_bound}
	still applies here.
\end{proof}

\begin{simplifying_assumption}\label{rule:capacity_lower_bound}
	$\capacity>2$.
\end{simplifying_assumption}
\begin{proof}
    Since all tiles contain at least one symbol,
    Rule~\ref{rule:tile_size_upper_bound} implies
    $\capacity\geq2$.
    Assume that $\capacity=2$,
    i.e., all tiles consist in one single symbol.
    Hence, $\{\{\tile_1,\tile_2\}, \{\tile_3,\tile_4\}, ...\}$
    is an optimal pagination of $\tileSet$
    in $\lceil{\frac{|\tileSet|}{2}\rceil}$ pages.
\end{proof}

\begin{simplifying_assumption}\label{rule:capacity_upper_bound}
	$\capacity<|\symbSet|$.
\end{simplifying_assumption}
\begin{proof} Otherwise, all tiles could fit in one page.
\end{proof}

To sum up,
an optimal solution of an instance violating
Rules~\ref{rule:no_inclusion},
\ref{rule:tile_size_upper_bound} (with $|\tile|=\capacity$),
\ref{rule:no_common_symbol},
\ref{rule:no_useless_symbol} or
\ref{rule:coexistence},
could be deduced from an optimal solution
of this instance deprived of the offending tiles or symbols;
an instance violating
Rule~\ref{rule:tile_size_upper_bound} (with $|\tile|>\capacity$)
would be infeasible;
an instance violating
Rules~\ref{rule:no_total_cover},
\ref{rule:capacity_lower_bound} or
\ref{rule:capacity_upper_bound}
would be trivial.
All these rules
can be tested in polynomial time,
and are actually required
by our instance generator
(Section~\ref{sec:test-set}).

\section{Heuristics}\label{sec:heuristics}

In this section, we investigate
four families of heuristics for \textsc{Pagination},
from the simplest to the most sophisticated one.
The first family consists
of direct adaptations
of the well-studied \textsc{Bin Packing}'s greedy \textsc{Any Fit} algorithms;
we show that,
in \textsc{Pagination},
their approximation factor cannot be bounded.
The second family is similar,
but
\replaced[id=R2]
    {relies}
    {rely}
on the overlapping property
specific to our problem;
a general instance is devised,
which shows that the approximation factor
of their offline version is at least 3.
With the third algorithm,
we leave the realm of greedy decisions
for a slightly more complex,
but hopefully more efficient strategy,
based upon a reentering queue.
Finally,
we present two genetic algorithms
and discuss which encoding and cost function
are better suitable to \textsc{Pagination}.
All of this is carried out from a theoretical perspective,
the next section being devoted
to the presentation of our benchmarks.

\subsection{Greedy heuristics inspired from \textsc{Bin Packing}}

\subsubsection{Definitions}

The question naturally arises of how the \textsc{Bin Packing} classical approximation algorithms \cite{Coffman:1996} behave in the more general case of \textsc{Pagination}.
Let us enumerate a few of them with our terminology:
\begin{itemize}
    \item \textsc{Next Fit} simply stores each new tile in the last created page or, should it exceed the capacity, in a newly created page.
    \item \textsc{First Fit} rescans sequentially the pages already created, and puts the new tile in the first page where it fits.
    \item \textsc{Best Fit} always chooses the fullest page, i.e., the page with maximal volume. Such a criterion needs to be clarified in our generalization of \textsc{Bin Packing}: fullest \emph{before} or \emph{after} having put the new tile? This alternative should give rise to two variants.
    \item \textsc{Worst Fit}, contrary to \textsc{Best Fit}, favors the less full page.
    \item \textsc{Almost Worst Fit} is a variant of \textsc{Worst Fit} opting for the \emph{second} less full page.
\end{itemize}

These algorithms are known under the collective name of \textsc{Any Fit} (AF).
In their offline version, pre-sorting the items by size has a positive impact on their packing; but for \textsc{Pagination}, such a sorting criterion would obviously be defective: due to possible merges, a large tile often occupies less volume than a small one.

\subsubsection{A general unfavorable case}\label{sec:af_greedy_bad_case}

Regardless of its scheduling (online or offline),
no AF algorithm has performance guarantee
on the following extensible \textsc{Pagination} instance.

\begin{figure}[htbp]
	\centering
	\includegraphics[width=\linewidth]{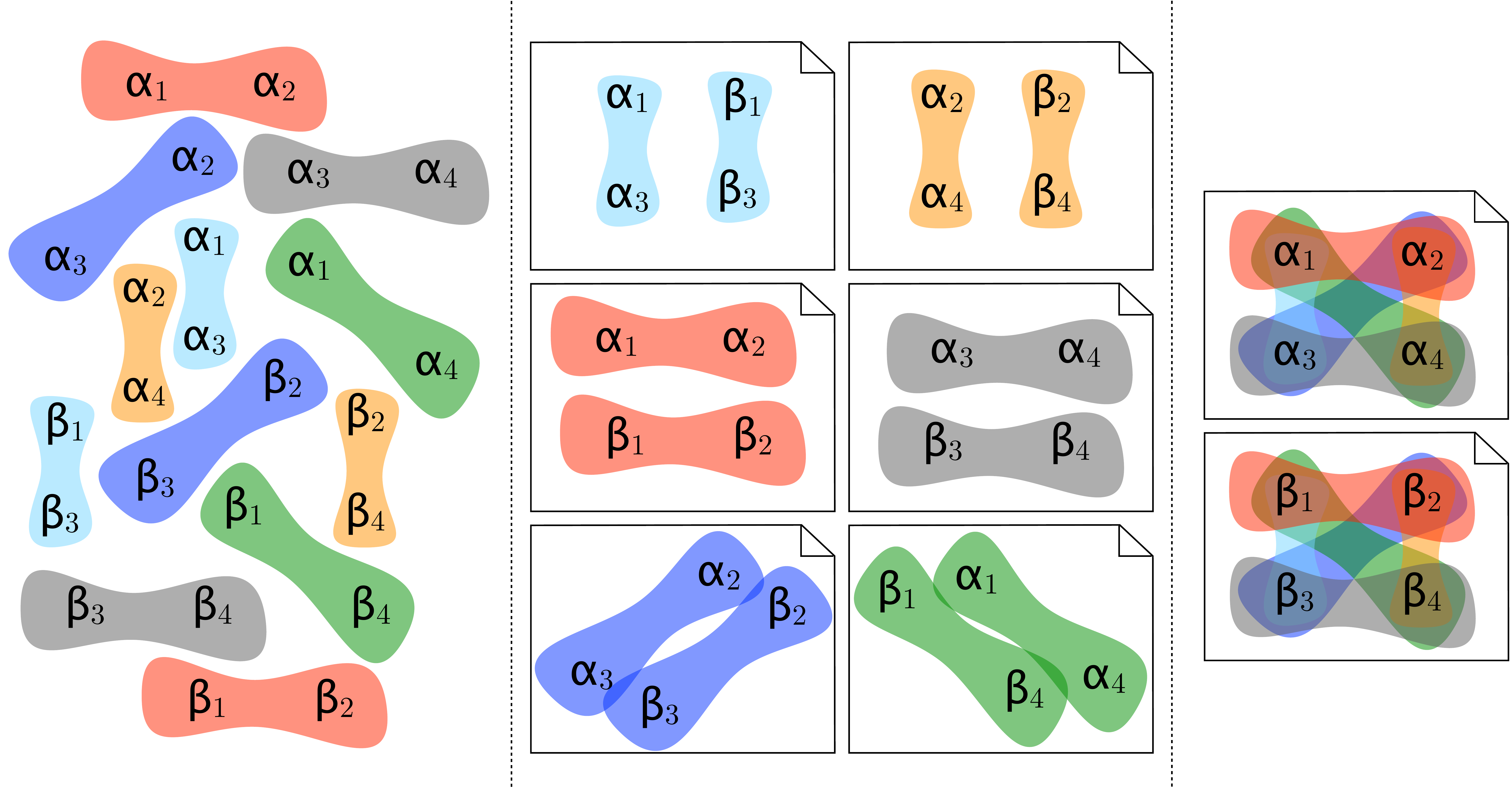}
	\caption{\label{fig:AF_worst_case_example}\it
    To the left, a set $\tileSet=\tileSet_\symbA\cup\tileSet_\symbB$ of $2\binom{4}{2}=12$ tiles; at the center, a pagination of capacity $4$ calculated by the \textsc{Any Fit} algorithms when the tiles of $\tileSet_\symbA$ and $\tileSet_\symbB$ are fed alternatively; to the right, an optimal pagination.}
\end{figure}

\added[id=R2]
    {Let $\capacity$ be an even capacity ($\capacity=4$ on Fig.~\ref{fig:AF_worst_case_example}).}
Let
$\symbSet_\symbA=\{\symbA_1,...,\symbA_\capacity\}$
and
$\symbSet_\symbB=\{\symbB_1,...,\symbB_\capacity\}$
be two
\replaced[id=R2]
    {sets of distinct symbols
    such that $\symbSet$ is the disjoint union of $\symbSet_\symbA$ and $\symbSet_\symbB$.}
    {disjoint subsets of size $\capacity$,
    assumed to be even
    ($\capacity=4$ on Fig.~\ref{fig:AF_worst_case_example}).}
Let
$\tileSet_\symbA=\binom{\symbSet_\symbA}{\capacity/2}$
and
$\tileSet_\symbB=\binom{\symbSet_\symbB}{\capacity/2}$
be the set of the $\frac{\capacity}{2}$-combinations of
$\symbSet_\symbA$ and $\symbSet_\symbB$ (respectively). Then $\pageSet_\textrm{opt}=\{\tileSet_\symbA,\tileSet_\symbB\}$ is an optimal pagination of $\tileSet=\tileSet_\symbA\cup\tileSet_\symbB$ in $2$ pages.

Now, let us feed these tiles to any of our AF algorithms,
but only after having sorted them in the worst order:
since all of our tiles have the same size $\capacity/2$,
we are indeed free to organize them the way we want.
The most unfavorable schedule simply involves alternating the tiles of $\tileSet_\symbA$ and $\tileSet_\symbB$. In this way,
regardless of the selected AF algorithm,
the first two tiles would saturate the first page, the next two, the second page, and so on.
In total, an AF algorithm would then create
$
\pageCount
	=|\tileSet_\symbA|
	=|\tileSet_\symbB|
	=\binom{\capacity}{\capacity/2}
	% =\frac{\capacity!}{(\capacity/2)!(\capacity-\capacity/2)!}
	=\frac{\capacity!}{(\capacity/2)!^2}
$
pages instead of 2.
In this family,
the approximation factor is thus unbounded.
This stands in stark contrast
to the efficiency of the AF algorithms on \textsc{Bin Packing}---where,
for instance, \textsc{First Fit} was recently shown \cite{Dosa:2013} to achieve
an absolute factor of exactly 1.7.

\subsubsection{Study case: First Fit algorithm}
For our tests,
we have chosen to focus
on \textsc{First Fit}
(in its online version).
The worst case complexity can be analyzed as follows.
There are $|\tileSet|$ tiles to paginate.
At worst,
according to Rule~\ref{rule:coexistence},
there exists only one tile compatible with any other,
resulting in $|\tileSet|-1$ pages.
Lastly,
each set intersection costs a linear time in the size of the candidate tile.
Hence,
the overall complexity is
$\mathcal{O}(|\tileSet|^2\card{\tileSet})$.

Note that in \textsc{Bin Packing},
for $n$ items,
an appropriate data structure can reduce
the straightforward quadratic complexity
to $\mathcal{O}(n\log(n))$ \cite{Johnson:1973}.
This optimization is not applicable here,
where the current volume of a given page
says little about its ability to accommodate a given tile.

\subsection{Specialized greedy heuristics}

\subsubsection{Definition}\label{sec:specialized-greedy-heuristics-definition}

We can easily improve
on the \textsc{Bin Packing} heuristics
by taking into account the merging property
of \textsc{Pagination} items.

The corresponding offline greedy algorithms
always select,
among the remaining tiles,
the one which minimizes or maximizes
a certain combination of the metrics
introduced in Definition~\ref{def:metrics}
(e.g., volume, multiplicity, relative size, etc.).
In their online version,
since the new tile is given,
this optimization relies only
on the past assignments.

\subsubsection{A general unfavorable case}\label{sec:specialized_greedy_bad_case}

Before introducing the particular heuristic
we have the most thoroughly tested,
\textsc{Best Fusion},
let us mark out the limitations
of the algorithms of this family.
We will design an extensible instance
whose all tiles are equivalent
with respect to these metrics.
As such,
they may be taken in any order,
including the worst,
which ensures that the reasoning
is valid for both online and offline scheduling.

Take an even capacity $\capacity$,
and $\symbSet_0=\{\symbA_1,...,\symbA_\capacity\}$
a first subset of $\symbSet$.
Let $\tileSet_0=\binom{\symbSet_0}{\capacity-1}$
the $(\capacity-1)$-combinations of $\symbSet_0$,
which amount to $\capacity$ tiles of size $\capacity-1$.
From now on,
to better understand the general construction,
we will illustrate each step on an example with $\capacity=4$:
\[
\symbSet_0 = \tilex{1234}, \qquad \tileSet_0   = \{\tilex{123}\,\tilex{124}\,\tilex{134}\,\tilex{234}\}
\]
Introduce two more symbols,
namely $\textsf{a}$ and $\textsf{b}$,
which will be used to lock the pages.
Partition $\tileSet$ into $\frac{\capacity}{2}$ couples of tiles:
${(\tile_1,\tile_2),(\tile_3,\tile_4),...,(\tile_{\capacity-1},\tile_\capacity)}$.
From each such couple,
form $\symbSet_i=(\tile_{2i-1}\cap\tile_{2i})\cup\{\textsf{a},\textsf{b}\}$
with $1 \leq i \leq \frac{\capacity}{2}$,
a subset of $\symbSet$.
In the same way as on $\symbSet_0$,
define $\tileSet_i=\binom{\symbSet_i}{\capacity-1}$ on each $\symbSet_i$:
\begin{align*}
\symbSet_1 &= \tilex{12ab}, \qquad \tileSet_1 = \{\tilex{12a}\,\tilex{12b}\,\tilex{1ab}\,\tilex{2ab}\} \\
\symbSet_2 &= \tilex{34ab}, \qquad \tileSet_2 = \{\tilex{34a}\,\tilex{34b}\,\tilex{3ab}\,\tilex{4ab}\}
\end{align*}
Constructing this instance is actually
constructing an optimal solution to it.
Indeed,
for the whole tile set,
$|\pageSet_\textrm{opt}|=\frac{\capacity}{2}+1$:
\begin{align*}
\page_1 &= \{\tilex{123}\,\tilex{124}\,\tilex{134}\,\tilex{234}\} \rightarrow \tilex{1234} \\
\page_2 &= \{\tilex{12a}\,\tilex{12b}\,\tilex{1ab}\,\tilex{2ab}\} \rightarrow \tilex{12ab} \\
\page_3 &= \{\tilex{34a}\,\tilex{34b}\,\tilex{3ab}\,\tilex{4ab}\} \rightarrow \tilex{34ab}
\end{align*}

Now, let us construct a non-optimal pagination.
This is done by pairing each tile of $\tileSet_0$
with a tile including a “locking” symbol.
Specifically,
on pages $2i-1$ and $2i$,
put respectively $\tile_{2i-1}$ and $\tile_{2i}$,
and lock these pages with tiles
$(\tile_{2i-1}\cap\tile_{2i})\cup\{\textsf{a}\}$ and  $(\tile_{2i-1}\cap\tile_{2i})\cup\{\textsf{b}\}$ 
(respectively)
from $\tileSet_i$.
This process creates $\capacity$ locked pages:
\begin{align*}
\page'_1 &= \{\tilex{123}\,\tilex{12a}\} \rightarrow \tilex{123a} \\
\page'_2 &= \{\tilex{124}\,\tilex{12b}\} \rightarrow \tilex{124b} \\
\page'_3 &= \{\tilex{134}\,\tilex{34a}\} \rightarrow \tilex{134a} \\
\page'_4 &= \{\tilex{234}\,\tilex{34b}\} \rightarrow \tilex{234b}
\end{align*}
But what enables us to construct such inefficient two-tile pages?
All tiles having the same size,
it is clear that the first one can be chosen arbitrarily.
Now, let $\tile'$ be an eligible second tile,
and $\page'$ the resulting page.
Then,
all $\tile'$ are equivalent under our various metrics:
same size $|\tile'|=\capacity-1$,
same volume $\volume{\page'}=\capacity$,
same cardinality $\card{\page'}=2(\capacity-1)$,
same relative size $\footprint{\page'}{\tile'}=\frac{\capacity-2}{2}+1=\frac{\capacity}{2}$.
So, nothing prevents
our greedy algorithms
to systematically select
the worst candidate.

If $\capacity>2$,
the $\capacity-2$ tiles
including both $\textsf{a}$ and $\textsf{b}$
still remain
in every tile set (but $\tileSet_0$).
For each one,
gather its tiles on a new page:
\begin{align*}
\page'_5 &= \{\tilex{1ab}\,\tilex{2ab}\} \rightarrow \tilex{12ab} \\
\page'_6 &= \{\tilex{3ab}\,\tilex{4ab}\} \rightarrow \tilex{34ab}
\end{align*}
Finally, we have obtained a non-optimal pagination $\pageSet$ totaling
$\capacity + \frac{\capacity}{2} = \frac{3\capacity}{2}$ pages.
Hence,
for a given even capacity $\capacity$,
any greedy algorithm of this family may yield
$\frac{|\pageSet|}{|\pageSet_\textrm{opt}|}=
\frac{3\capacity}{\capacity+2}$ times more pages than the optimal.
In other words,
its approximation factor is at least 3.

\subsubsection{Study case: Best Fusion algorithm}

In our benchmarks,
the following criterion was used:
for each tile $\tile$,
let $\page$ be the eligible page
on which the relative size $\footprint{\page}{\tile}$ is minimal.
If $\footprint{\page}{\tile}<|\tile|$, then put $\tile$ on $\page$;
otherwise, put $\tile$ on a new page.
We call \textsc{Best Fusion}
the online version of this algorithm.
Its worst-case complexity is the same as in \textsc{First Fit},
i.e.,
$\mathcal{O}(|\tileSet|^2\card{\tileSet})$.

Since a new page is created for every tile
whose assignment to an existing page
would bring no immediate benefit,
the cases raised in Section~\ref{sec:af_greedy_bad_case}
are solved to optimality.
However,
the downside of such a strategy
becomes readily apparent
with pure \textsc{Bin Packing} instances:
in the absence of shared symbols,
each new tile will trigger the creation of a new page.
Consequently,
\textsc{Best Fusion} has no more performance guarantee
than the AF algorithms.
The difference here is that
no resulting page is locked in a bad state.
In practice,
this flaw is easily fixed
by the so-called decantation post-treatment
(see Section~\ref{sec:decantation}),
which can be considered
as a multi-scale \textsc{First Fit}%
\footnote{
Strictly speaking,
with this post-treat\deleted[id=R2]{e}ment,
\textsc{Best Fusion} is no more greedy.
Preserving at the same time
the greediness
of the algorithms of the present family,
and an acceptable behavior on both
Section~\ref{sec:af_greedy_bad_case}'s
and pure \textsc{Bin Packing}'s instances,
can nevertheless be attained
by means of offline scheduling
(described in Section~\ref{sec:specialized-greedy-heuristics-definition}).
This is at least $|\tileSet|$ times more complex, and still untested.
}.

\subsection{Overload-and-Remove heuristic}

The following non-greedy approach
has the ability
to reconsider past choices
whenever better opportunities arise.
As a result,
in particular,
it always finds the optimal solution
of the unfavorable cases
outlined in Sections~\ref{sec:af_greedy_bad_case} and \ref{sec:specialized_greedy_bad_case}.

The main idea is to add a given tile $\tile$
to the page $\page$
on which $\tile$ has the minimal relative size,
even if this addition actually overloads $\page$.
In this case,
the algorithm immediately tries to unload $\page$
by removing the tile(s) $\tile'$ of strictly \replaced[id=R2]
    {smallest}
    {smaller}
$\text{size}\over\text{relative size}$ ratio.
The removed tiles are rescheduled at the latest
by adding them to a FIFO data structure,
and simultaneously forbidden to reenter the same page---this ensures termination.
When the main loop is over,
the possible remaining overloaded pages are suppressed,
and their tiles redistributed by \textsc{First Fit}.

\begin{algo}
\newcommand{\tab}{|\hspace{1em}}
% \colorbox{lightgray}{\pbox{0.97\columnwidth}{
\sf
\textrm{Algorithm: \textsc{Overload-and-Remove}}\\
\rule[1ex]{\linewidth}{0.4pt}\\
$\queue\leftarrow$ queue containing all the tiles of $\tileSet$\\
$\pageSet\leftarrow$ pagination consisting of one empty page\\
\textbf{while} $\queue$ is nonempty:\\
\tab{}$\tile\leftarrow$ dequeue($\queue$)\\
\tab{}$\pageSet_\tile\leftarrow$ pages of $\pageSet$ where $\tile$ has never been put on\\
\tab{}\textbf{if} $\pageSet_\tile$ has no page $\page$ such that $\footprint{\page}{\tile}<|\tile|$:\\
\tab{}\tab{}add to $\pageSet$ a new page consisting solely of $\{\tile\}$\\
\tab{}\tab{}\textbf{continue with next iteration}\\
\tab{}$\page\leftarrow$ page $\page$ of $\pageSet_\tile$ such that $\footprint{\page}{\tile}$ is minimal\\
\tab{}put tile $\tile$ on page $\page$\\
\tab{}\textbf{while} $\volume{\page} > \capacity$
    \textbf{and}
    $
    % \min_{\tile\in\page}{\frac{|\tile|}{\footprint{\page}{\tile}}}
    % <
    % \max_{\tile\in\page}{\frac{|\tile|}{\footprint{\page}{\tile}}}
    \exists \tile_1,\tile_2\in\page^2\suchthat
    \frac{|\tile_1|}{\footprint{\page}{\tile_1}}
    \neq
    \frac{|\tile_2|}{\footprint{\page}{\tile_2}}
    $:\\
\tab{}\tab{}remove from $\page$ one tile $\tile'$
    minimizing $|\tile'|/\footprint{\page}{\tile'}$\\
\tab{}\tab{}enqueue($\queue, \tile'$)\\
remove all the overloaded pages from $\pageSet$\\
put their tiles back in $\pageSet$ (by First Fit)
% }}
\end{algo}

In the worst case,
a given tile might successively overload and be removed
from all pages,
whose total number is at most $|\tileSet|$.
Each trial requires one set intersection
on each page.
Hence, the overall complexity is
$\mathcal{O}(|\tileSet|^3\card{\tileSet})$.

\subsection{Genetic algorithms}
\subsubsection{Standard model}

\paragraph{Encoding}

Any pagination (valid or not) on $\pageCount$ pages is encoded as a tuple
$(\pageIndex_1,...,\pageIndex_{|\tileIndexSet|})$
where $\pageIndex_\tileIndex\in[1,\pageCount]$
is the index of the page containing the tile $\tile_\tileIndex$.

\begin{example}
	The four paginations of Fig.~\ref{fig:intro} would be respectively encoded as $(1,2,3,4)$, $(1,2,2,3)$, $(1,1,2,2)$ and $(1,1,1,2)$.
\end{example}

Due to its overly broad encoding capabilities,
\textsc{Standard GA} is \emph{not} guaranteed
to produce a valid pagination.
Our fitness function is
devised with this in mind,
in such a way that an invalid chromosome
would always cost more than a valid one.
Thus,
seeding the initial population
with at least one valid individual
will be enough to ensure success.

\paragraph{Evaluation}
Our aim is twofold.
First and foremost, 
to penalize the invalid paginations;
second,
to reduce the volume of the last nonempty page.
For this purpose,
we will minimize the fitness function $\fitness$ defined as follows:
\begin{subequations}

As soon as one page is overloaded
(i.e., $\exists\pageIndex\in\pageIndexSet\suchthat
\sum_{\symbIndex\in\symbIndexSet}\symbInPage{\symbIndex}{\pageIndex}>\capacity$),
we count $|\tileSet|\capacity$ symbols (as if all possible pages were saturated),
to which we also add every extra symbol:
\begin{equation}\label{eq:standard:fitness:invalid}
\fitness(\pageSet) =
|\tileSet|\capacity+\sum_{\pageIndex\in\pageIndexSet}
{\max\big(0,(
\sum_{\symbIndex\in\symbIndexSet}\symbInPage{\symbIndex}{\pageIndex})-\capacity\big
)}
\end{equation}

Otherwise,
let us call $\pageIndex$ the index of the last nonempty page
(i.e., such that $\forall \pageIndex'>\pageIndex, \pageUsed{\pageIndex'}=0$).
Count $(\pageIndex-1)\capacity$ symbols
(as if all nonempty pages but the last one were saturated),
and add the number of symbols on page $\page_\pageIndex$:
\begin{equation}\label{eq:standard:fitness:valid}
\fitness(\pageSet) = (\pageIndex-1)\capacity+\displaystyle\sum_{\symbIndex\in\symbIndexSet}\symbInPage{\symbIndex}{\pageIndex}
\end{equation}
\end{subequations}

\begin{example}
    If all tiles of Fig.~\ref{fig:intro} were put on a single (invalid) page,
    by (\ref{eq:standard:fitness:invalid}),
    the fitness value would reach $4\times7+4=32$.
    By (\ref{eq:standard:fitness:valid}),
    the paginations $\pageSet_1$ to $\pageSet_4$ have a fitness value of $3\times7+4=25$, $2\times7+4=18$, $1\times7+7=14$, and $1\times7+4=11$ (respectively).
\end{example}

\paragraph{Mutation}
It consists in transferring one randomly selected tile from one page to another.

\paragraph{Crossover}
The standard two-point crossover applies here without requiring any repair process.

% First of all,
% we should point out that
% \textsc{Standard GA} produced no valid pagination
% on 605 instances\footnote{
% Regarding \Fig.~\ref{fig:relative_size_by_multiplicity},
% which uses a rolling window of width 150,
% please note that a single missing data point would normally
% prevent any calculation on its $2 \times 149$ nearest neighbors,
% which would make the \textsc{Standard GA} line almost disappear.
% To avoid this outcome,
% we have chosen to replace each missing result
% by the worst result given by the other algorithms on the same instance.
% The resulting line is connected,
% but remember this is a quite optimistic view
% of its actual behavior.
% }.
% Although this constitutes only 5.51~\% of cases,
% the failure rate increases sharply with the \added[id=R2]{statistical} difficulty:
% already at 6~\% when multiplicity passes 3.6,
% it reaches 50~\% at 28.5,
% and 90~\% above 48.5.
% Not content to be prone to failure,

\subsubsection{Grouping model}\label{sec:grouping_model}

In \cite{Falkenauer:1992},
Falkenauer et al. show that classic GAs are not suitable
to the \term{grouping problems}, namely
“optimization problems where the aim is
to group members of a set into a small number of families,
in order to optimize a cost function,
while complying with some hard constraints”.
To take into account the structural properties of such problems,
the authors introduce the so-called \term{grouping genetic algorithms} (GGA).
Their main idea is to encode each chromosome
on a one gene for one group basis.
The length of these chromosomes is thus variable:
it depends on the number of groups.
Crucially,
the belonging of several items to \replaced[id=R2]{the same}{a same} group
is protected during crossovers: the good schemata
are more likely to be transmitted to the next generations.

\textsc{Pagination} is clearly a grouping problem,
moreover directly derived from \textsc{Bin Packing}---one of the very problems Falkenauer
chooses to illustrate his meta-heuristic.
We thus will adapt,
and sometimes directly apply his modelization.

\paragraph{Encoding}

A \emph{valid} pagination on $\pageCount$ pages is encoded as a tuple
$(\page_1,...,\page_\pageCount)$
where $\page_\pageIndex$ is the set of the indexes
of the tiles put on the $\pageIndex^{\text{th}}$ page.

\begin{example}
	This time, the paginations of Fig.~\ref{fig:intro} would be encoded as
$(\{1\},\allowbreak\{2\},\allowbreak\{3\},\allowbreak\{4\})$,
$(\{1\},\allowbreak\{2,3\},\allowbreak\{4\})$,
$(\{1,2\},\allowbreak\{3,4\})$
and
$(\{1,2,3\},\allowbreak\{4\})$
(respectively).
Or,
in Falkenauer's indirect, but sequential notation:
1234:1234, 1223:123, 1122:12 and 1112:12,
where the left part denotes,
    for each $\tileIndex^\text{th}$ tile,
    the index of its page;
and the right part,
    the list of all pages.
\end{example}

\paragraph{Evaluation}

In our notation,
the maximization function of \cite{Falkenauer:1992} for \textsc{Bin Packing}
would be expressed as
$
\fitness_\textrm{BP}(\pageSet) = \frac{1}{\pageCount}
\sum_{\pageIndex=1}^{\pageCount}
(\frac{1}{\capacity}\volume{\page_\pageIndex})^\disparity.
$
In other words,
the average of volume rates
raised to a certain \textbf{disparity} $\disparity$,
which sets the preference given to the bins' imbalance:
thus,
for \replaced[id=R2]{the same}{a same} number of bins and \replaced[id=R2]{the same}{a same} total loss,
the greater the disparity,
the greater the value of an unbalanced packing.

Although this formula still makes sense in the context of \textsc{Pagination},
we should not apply it as is.
Indeed,
whereas minimizing the loss
amounts to minimizing the number of bins,
Proposition~\ref{prop:min_loss_not_opt} warns us
this is actually untrue for the number of pages:
in the associated counterexample,
with $\disparity=2$ (the empirical value proposed by \cite{Falkenauer:1992}),
the optimal pagination would be evaluated to
$((3/4)^2+(3/4)^2)/2=0.5625$,
and the suboptimal one to
$((4/4)^2+(4/4)^2+(4/4)^2)/3=1$.

Instead of privileging the high volume pages,
we will privilege the high multiplicity ones
(i.e., replace $\volume{\page_\pageIndex}$
by $\card{\page_\pageIndex}$).
Proposition~\ref{prop:opt_eq_max_avg_weight} guarantees
that the higher the average page multiplicity,
the better the overall pagination.

One detail remains to be settled:
ensure that the quantity
raised to the power $\disparity$ never exceeds 1.
Here, in the same way that
$\volume{\page_\pageIndex}$ is bounded by $\capacity$,
$\card{\page_\pageIndex}$ is bounded by $\card{\tileSet}$,
\replaced[id=R2]
    {and even, more tightly,
    by the sum of the multiplicities of the $\capacity$ most common symbols,
    which we will note $M_\capacity^\tileSet$.
    This leads us to}
    {Although a tighter bound can be, and has been implemented
    (namely, the sum of the multiplicities of the $\capacity$ most common symbols),
    to make it short we will halt on}
the following fitness function for \textsc{Pagination}:

\begin{equation}\label{eq:fitness-group}
\fitness(\pageSet) = \frac{1}{\pageCount}
\sum_{\pageIndex=1}^{\pageCount}
\Big(\frac{\card{\page_\pageIndex}}
    {M_\capacity^\tileSet}
    \Big)^\disparity
\end{equation}

\begin{example}
    In Fig.~\ref{fig:intro},
    \replaced[id=R2]
        {$M_\capacity^\tileSet = 3+2+2+1+1+1+1 = 11$.}
        {$\card{\tileSet} = 5+3+3+4 = 15$.}
    With $\disparity=2$,
    the four paginations
    are respectively evaluated as follows:
    \begin{align*}
		\left((5/11)^2+(3/11)^2+(3/11)^2+(4/11)^2\right)/4&\simeq0.12\\
		\left((5/11)^2+(6/11)^2+(4/11)^2\right)/3&\simeq0.21\\
		\left((8/11)^2+(7/11)^2\right)/2&\simeq0.47\\
		\left((11/11)^2+(4/11)^2\right)/2&\simeq0.57
	\end{align*}
    As one can see,
    $\pageSet_4$ (the most unbalanced pagination) scores better than $\pageSet_3$.
    The difference would increase with disparity $\disparity$.
\end{example}

\paragraph{Mutation}
The mutation operator of \cite{Falkenauer:1992}
consists in emptying at random a few bins,
shuffling their items,
and then inserting them back by \textsc{First Fit}.
We follow the exact same procedure,
but without the suggested improvements:
at least three reinserted bins,
\replaced[id=R2]
    {among which}
    {of whom}
the emptiest one.

\paragraph{Crossover}
The two parents (possibly of different length)
are first sliced into three segments:
$(a_1,a_2,a_3)$ et $(b_1,b_2,b_3)$.
The tiles of $b_2$ are withdrawn from $a_1$ and $a_3$:
$a_1'=a_1\setminus b_2$ and $a_3'=a_3\setminus b_2$.
To construct the first offspring,
we concatenate $(a_1',b_2,a_3')$,
sort the missing tiles in decreasing order,
and insert then back by \textsc{First Fit}.
The second offspring is calculated in the same manner
(just exchange the roles of $a$ and $b$).

\subsection{Post-treatment by decantation}\label{sec:decantation}

We introduce here a quadratic algorithm
which,
in an attempt to reduce the number of pages,
will be systematically applied
to the results produced
by all our heuristics but \textsc{First Fit}.
Its three steps consist in settling
at the beginning of the pagination
as much pages, components and tiles as possible.
First, we must specify what is a component:

\begin{definition}
    Two tiles are \term{connected} if and only if
    they share at least one symbol
    or \replaced[id=R2]
        {are both connected to the same intermediate tile.}
        {are connected to another tile.}
    The \term{(connected) components} are the classes
    of the associated equivalence relation.
\end{definition}

\begin{example}
  In Fig.~\ref{fig:intro},
  the components of the instance 
are $\{\tilex{abcde}\,\tilex{def}\,\tilex{efg}\}$
and $\{\tilex{hijk}\}$.
\end{example}

\begin{definition}
A valid pagination is said to be \term{decanted} on the pages
(resp., components, tiles)
if and only if no page contents
(resp., component, tile)
can be moved to a page of lesser index
without making the pagination invalid.
\end{definition}

\begin{example}
In Fig.~\ref{fig:intro},
$\pageSet_3$ is decanted on the pages,
but not on the components or the tiles.
$\pageSet_4$
is a fully decanted pagination (on the pages, the components and the tiles).
\end{example}

Obviously,
a pagination decanted on the tiles
is decanted on the components;
and a pagination decanted on the components
is decanted on the pages.
To avoid any unnecessary repetition,
the corresponding operations
must then be carried out
in the reverse direction. Moreover,
the best decantation of a given pagination $\pageSet$
is attained by decanting $\pageSet$
successively
on the pages, the components, and then the tiles.

\begin{example}
Let $(
\{\tilex{123}\},\allowbreak
\{\tilex{14}\,\tilex{567}\},\allowbreak
\{\tilex{189}\}
)$ 
be a pagination in 3 pages with $\capacity=5$.
Its decantation on the components, $(
\{\tilex{123}\,\tilex{14}\},\allowbreak
\{\tilex{567}\},\allowbreak
\{\tilex{189}\}
)$,
does not decrease the number of pages,
as opposed to its decantation on the pages: $(
\{\tilex{123}\,\tilex{189}\},\allowbreak
\{\tilex{14}\,\tilex{567}\}
)$.
For an example on components/tiles,
\added[id=R2]
    {let us}
substitute $\tilex{567}$ with $\tilex{1567}$
in the instance.
\added[id=R2]
    {Let $(
    \{\tilex{123}\},\allowbreak
    \{\tilex{14}\,\tilex{1567}\},\allowbreak
    \{\tilex{189}\}
    )$ 
    be a pagination in 3 pages with $\capacity=5$.
    Its decantation on the tiles, $(
    \{\tilex{123}\,\tilex{14}\},\allowbreak
    \{\tilex{1567}\},\allowbreak
    \{\tilex{189}\}
    )$,
    does not decrease the number of pages,
    as opposed to its decantation on the components: $(
    \{\tilex{123}\,\tilex{189}\},\allowbreak
    \{\tilex{14}\,\tilex{1567}\}
    )$.}
\end{example}

This decantation algorithm is thus implemented
as a sequence of three \textsc{First Fit} procedures
of decreasing granularity.
In our tests,
\replaced[id=R2]
    {the interest of such a post-treatment varies greatly:
    it is of course useless on \textsc{First Fit},
    but reduces the outcomes of GAs by one page in about 0.1~\% of cases,
    and both \textsc{Best Fusion} and \textsc{Overload-and-Remove}
    by up to ten pages for about 21~\% and 40~\% of instances,
    respectively.}
    {for any heuristic except \textsc{First Fit} itself,
    it was not unusual to gain one page,
    and even two,
    on the final pagination.}

\section{Experimental results}\label{sec:experimental-results}

\paragraph{Supplementary material}
In order to empower the interested readers
to reproduce our analysis
and conduct their own investigation,
we provide at \cite{notebook:2017}
a $\raise.17ex\hbox{$\scriptstyle\sim$}$60~MB Git repository
containing:
the whole set of our random instances (\texttt{gauss});
a companion Jupyter Notebook (\texttt{analysis.ipynb})
which generates
every plot and numerical result
mentioned or alluded in the present section;
some instructions
for using this notebook interactively
(\texttt{README.md}).

\subsection{Generating a test set}\label{sec:test-set}

First, let us introduce another simplifying assumption:

\begin{simplifying_assumption}[optional]\label{rule:tile_size_lower_bound}
	All tiles contain more than one symbol: $\forall\tile\in\tileSet,|\tile|>1$.
\end{simplifying_assumption}

This rule cannot be considered
on the same level than those of Section~\ref{sec:rules}:
forbidding one-symbol tiles
may theoretically make some instances
easier to paginate.

\begin{proof}[counterexample]
    By definition, no tile is empty.
    Suppose there exists a tile $\tile$ of size 1,
    and let $\pageSet'$ be an optimal pagination
    of the reduced instance $\tileSet\setminus\{\tile\}$.
    If there exists a page $\page'\in\pageSet'$ such that $|\page'|<\capacity$,
    then adding $\tile$ on $\page'$
    produces an optimal pagination of $\tileSet$.
    However,
    in the rare cases where all pages are saturated,
    $\pageSet'\cup\{\{\tile\}\}$
    is not necessarily
    an optimal pagination of $\tileSet$.
    Take for instance
    $\tileSet=\{\tilex{0}\,\allowbreak\tilex{123}\,\allowbreak\tilex{45}\,\allowbreak\tilex{167}\,\allowbreak\tilex{89}\}$
    and $\capacity=5$.
    The reduced instance $\tileSet\setminus\{\tilex{0}\}$
    admits an optimal pagination
    $\pageSet'=\{\{\tilex{123}\,\allowbreak\tilex{45}\}, \{\tilex{167}\,\allowbreak\tilex{89}\}\}$\replaced[id=R2]
        {, all of whose pages}
        {whose all pages}
    are saturated.
    Nevertheless,
    the pagination $\pageSet'\cup\{\{\tilex{0}\}\}$
    has one page more than the optimal pagination
    $\pageSet=\{\{\tilex{123}\,\allowbreak\tilex{167}\}, \{\tilex{45}\,\allowbreak\tilex{89}\,\allowbreak\tilex{0}\}\}$.
\end{proof}

Since the latter behavior requires both 
a specially constructed instance
\emph{and}
a particularly weak pagination algorithm,
we have chosen to add this constraint
during the generation of our random test sets:
in practice,
it will have no other effect than
to avoid polluting the instances with
a bunch of unmergeable, interchangeable, easy-to-place tiles.

\bigskip

Our instance generator takes as input
a capacity $\capacity$,
a number $|\symbSet|$ of symbols
and
a number $|\tileSet|$ of tiles.
\added[id=R2]{It follows a three-step process:}
\begin{enumerate}
\item \added[id=R2]{Calculate 
    a standard deviation $s=\frac{\capacity}{5}$
    and
    a mean $m=\frac{R\capacity}{4}$,
    where $R$ is a random variable uniformly distributed on $\{1,2,3\}$.}
\item \added[id=R2]{Draw an integer $k$ from the corresponding normal distribution:
    $k \sim \mathcal{N}(m, s^2)$.
    If $1<k<\capacity$ (Rules~\ref{rule:tile_size_upper_bound} and \ref{rule:tile_size_lower_bound}),
    make a candidate tile up from
    a uniform random sample of $k$ distinct symbols.
    Add it to the accepted set if and only if
    Rule~\ref{rule:no_inclusion}
    is still satisfied.
    Repeat this step until $|\tileSet|$ tiles are obtained.}
\item \added[id=R2]{Accept the resulting instance if and only if
    Rules~\ref{rule:no_common_symbol},
    \ref{rule:no_useless_symbol} and \ref{rule:coexistence} are satisfied.}
\end{enumerate}
\deleted[id=R2]{
First, an integer $k$ is drawn from a normal distribution
(whose mean and standard deviation depend on $C$ and a uniform random factor).
If $1<k<|\symbSet|$ (Rules~\ref{rule:no_total_cover} and \ref{rule:tile_size_lower_bound}),
a candidate tile is made up from
a uniform random sample of $k$ distinct symbols.
It is added to the accepted set if and only if
Rules~\ref{rule:no_inclusion} and
\ref{rule:tile_size_upper_bound}
are still satisfied.
The process is repeated until $|\tileSet|$ tiles are obtained.
To be accepted, the resulting instance must finally satisfy
Rules~\ref{rule:no_common_symbol},
\ref{rule:no_useless_symbol} and \ref{rule:coexistence}.
}
This algorithm has been called repeatedly with
$\capacity$ varying from $15$ to $50$ in steps of $5$,
$|\symbSet|$ varying from $\capacity + 5$ to $100$ in steps of $5$ (for lesser values, by Rule~\ref{rule:capacity_lower_bound}, all symbols would fit in a single page)
and $|\tileSet|$ varying from 
$20$ to $100$ in steps of $5$.
Although in some rare cases,
numerous passes were required
before halting on a result,
it proved to be robust enough to produce,
for each distinct combination of its parameters,
six valid instances
(for a total of 10,986 instances).

\subsection{Measuring the \added[id=R2]{statistical} difficulty of a given instance}

What is a \emph{difficult} instance of \textsc{Pagination}?
\replaced[id=R2]
    {Although we cannot answer this question in all generality,
    the highly experimental nature of the present approach
    certainly enables some observations to be made.}
    {Our proposed answer takes advantage of
    the highly experimental nature of the present approach.}
Indeed, we have not only generated several thousands of instances,
but also submitted them to
no less than six different solvers:
one ILP,
two genetic algorithms,
two greedy algorithms,
and one specialized heuristic (not counting its sorted variant),
\textsc{Overload-and-Remove}.
When these various methods all produce roughly the same number of pages,
one can conclude that the instance was an easy one;
conversely, when the pagination size differs greatly among methods,
it clearly presents some particular challenges.

The dispersion of the pagination sizes can be measured in several ways:
range (i.e., difference between minimum and maximum),
\added[id=R2]
    {average range (i.e., average difference with the minimum),}
standard deviation,
median absolute deviation...
Although, statistically speaking, the \replaced[id=R2]
    {last one}
    {latter}
is the most robust,
the \replaced[id=R2]{second one}{former}
proved to be slightly more suited to our problem,
where outliers are arguably no accidents,
but rather evidences of some significant structural features.
Hence:

\begin{conjecture}
    The \textbf{\added[id=R2]{statistical} difficulty} of a given instance
    can be approximated by the difference between
    the \replaced[id=R2]{average}{maximal} and the minimal number of pages
    in the paginations calculated by \replaced[id=R2]{our}{the} various solvers.
\end{conjecture}

There are some caveats.
First, this measure of \added[id=R2]{the statistical} difficulty
is intrinsically correlated (Pearson's \replaced[id=R2]{$r=0.792$}{$r=0.777$} \cite{Pearson:1895})
to the size of the best pagination
(e.g., if the best algorithm produces 2 pages,
it is unlikely that any of its competitors will produce 20 pages).
This is by design.
Intuitively, a “large” random instance is more difficult to paginate than a “small” one.
\added[id=R2]{For example, the larger an instance,
the longer it takes for a brute force algorithm to find the optimum.}
\replaced[id=R2]{In our tests}{Therefore},
normalizing the difference by dividing it by the best size \replaced[id=R2]{has indeed proven}{would be} counterproductive.

Second caveat, this measure only makes sense for random instances.
We certainly could devise a particular instance which would be easy for a special tailored algorithm,
but difficult for our general-purpose solvers.

\added[id=Authors]{Third caveat,
CPLEX gave us the optimal pagination for only 43 instances.
That the minimum heuristic result is the optimum,
or a tight estimate,
is therefore not guaranteed.
Note however that,
when the optimum is known,
the same number of pages is produced
by at least one heuristic in $1-4/43=91$~\% of the cases.
}

Finally,
\added[id=R2]{as its name suggests,}
this measure depends on our set of algorithms.
\added[id=R2]{Therefore, }
adding another one may \replaced[id=R2]
    {either increase or decrease the statistical difficulty of any instance.}
    {paradoxically increase the measured statistical difficulty of some instances.}
Of course, the more \added[id=R2]
    {“reasonable”}
algorithms would be tested,
the less the prevalence of this effect.

\subsection{Predicting the \added[id=R2]{statistical} difficulty of a given instance}\label{sec:avg-multiplicity}

\textsc{Pagination} can be seen as
an almost continuous extension of \textsc{Bin Packing}:
being given a pure \textsc{Bin Packing} instance
(i.e., no tile has shared symbols),
we may gradually increase its intricacy
by transforming it into
a \textsc{Pagination} instance as convoluted as desired
(i.e., many tiles share many symbols).
Therefore, we can expect that:

\begin{conjecture}
The \added[id=R2]{statistical} difficulty of a given random instance
is strongly correlated to the density of its shared symbols,
or \textbf{average multiplicity}.
\end{conjecture}

\begin{figure}[htbp]
\begin{center}
    \includegraphics[width=\linewidth]{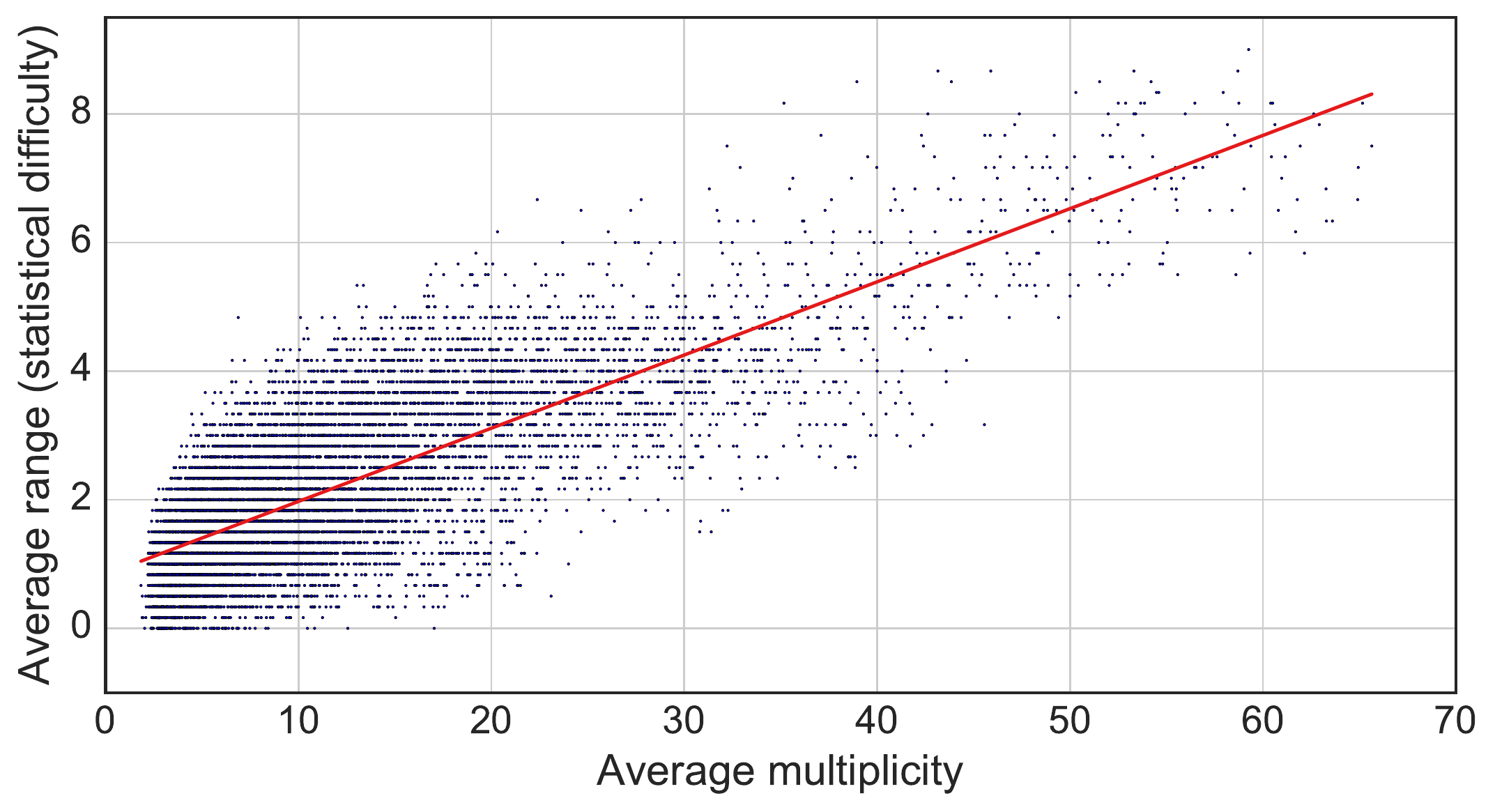}
    \caption{\label{fig:difficulty_by_multiplicity}\it
    \replaced[id=R2]{Statistical}{Average} difficulty by average multiplicity
    \added[id=Authors]{($r=0.784$)}.
    A multiplicity of 10 indicates that,
    in the corresponding instances,
    a given symbol is shared by an average of 10 tiles.
    For these instances,
    the average range of pagination sizes produced by our different solvers
    (the so-called difficulty)
    \replaced[id=Authors]
        {varies between 0 (all solvers obtain the same number of pages) and 5.}
        {is almost 4.}
    }
\end{center}
\end{figure}

\added[id=Authors]{
Figure~\ref{fig:difficulty_by_multiplicity}
tests this hypothesis on our instance set:
the average multiplicity of a given random instance
appears to be a good predictor ($r=0.784$)
of its statistical difficulty.}\added[id=R2]{
    Is it the best one?
    We can think of a few other possible candidates,
    and calculate their correlation with the statistical difficulty:}
\begin{itemize}
\item \added[id=R2]{$r=-0.082$ for $|\symbSet|$, the number of symbols;}
\item \added[id=R2]{$r=0.366$ for $|\symbSet|\times|\tileSet|$, the size of the instance,
    i.e., the number of bits required to encode it.
    More precisely,
    a stream of $|\symbSet|\times|\tileSet|$ bits,
    where the $i$th bit is 1 if and only if
    the $\lfloor i / |\tileSet|\rfloor$-th tile of $\tileSet$
    includes the $(i \mod |\tileSet|)$-th symbol of $\symbSet$
    (plus three long integers for
    the capacity,
    the size of the alphabet and
    the number of tiles,
    all of them being dominated by the first quantity).}
\item \added[id=R2]{$r=0.563$ for $|\tileSet|$, the number of tiles.}
\item \added[id=R2]{$r=0.770$ for $\card{\tileSet}$, the sum of the tile sizes.}
\end{itemize}
\added[id=R2]{
    Among these candidates, $\card{\tileSet}$, 
    the cardinality of the tile set,
    is almost as good a predictor as 
    $\card{\tileSet}/|\symbSet|$, its average multiplicity.
    The choice of the latter value is mainly dictated by the fact
    that it involves the number of symbols.
    Indeed,
    when $|\symbSet|$ approaches $\card{\tileSet}$,
    \textsc{Pagination} approaches \textsc{Bin Packing},
    which we expect to reduce the difficulty of the instance.
}

\begin{figure}[htbp]
\begin{center}
    \includegraphics[width=\linewidth]{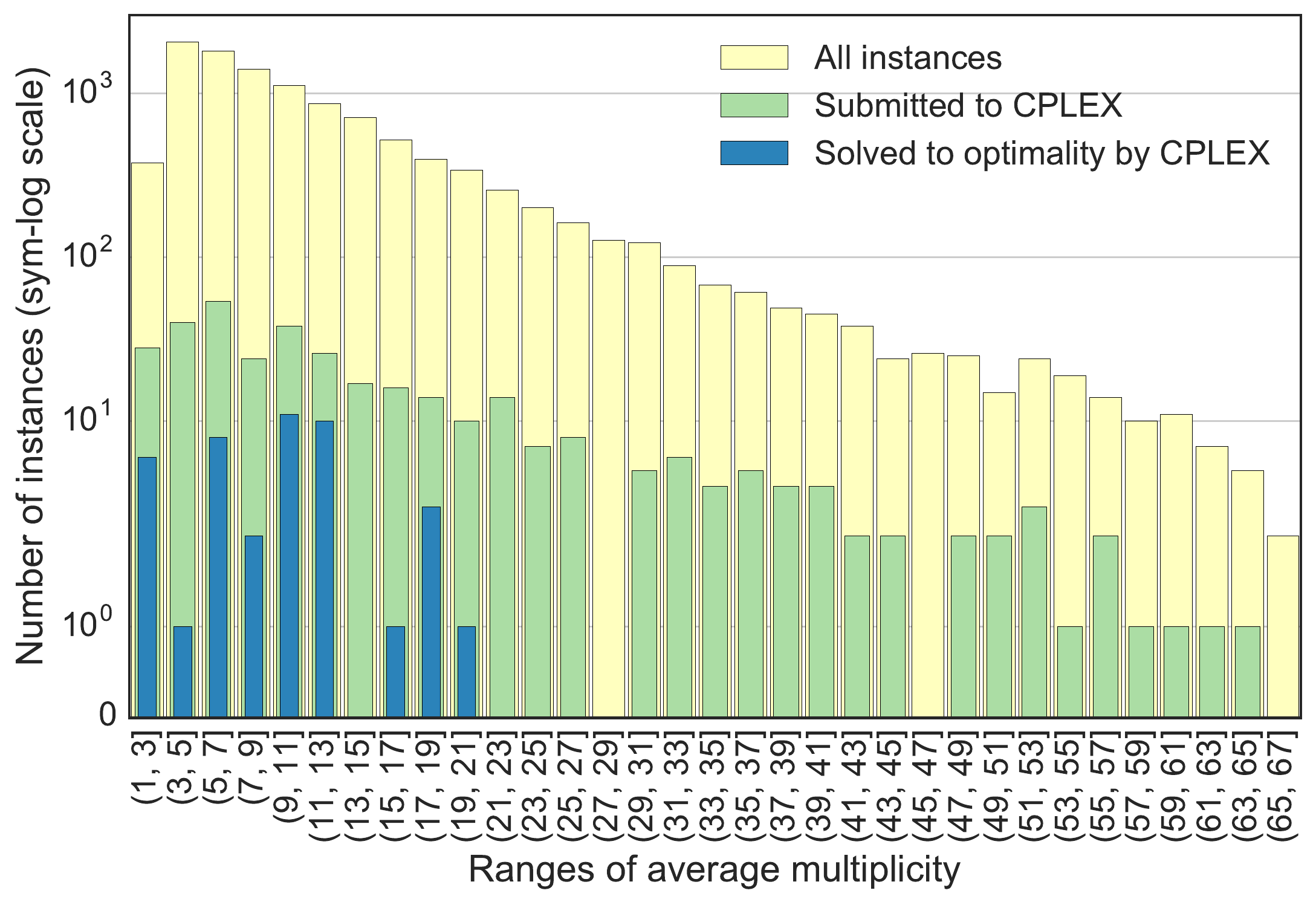}
    \caption{\label{fig:count_by_multiplicity}\it
    Number of instances by average multiplicity
    (sym-log scale).
    }
\end{center}
\end{figure}

\added[id=Authors]{
Before we go any further,
it is important to be aware
that the average multiplicities in our test set
are far from being evenly distributed
(Fig.~\ref{fig:count_by_multiplicity}):
for example,
there are 1119 instances whose \added[id=Authors]{average} multiplicity lies between 4 and 5,
but only 107 between 23 and 24,
and 10 between 53 and 54
(see \cite{notebook:2017} for these exact values).
Overall,
more than half of them
concentrate between multiplicities 2 and 9.
Thus, any observation made
on the higher multiplicities (and the smaller ones)
must be approached with great caution.}

\deleted[id=Authors]{
    To take this into account,
    we carry out our analysis
    on a moving window
    of equally-sized subsets of instances
    sorted by increasing average multiplicity:
    the disappearance of some endpoints
    is greatly outweighed
    by the gain
    of a constant confidence level
    on the remaining data.
    And indeed,
    on our extensive test set,
    the average multiplicity of a given random instance
    appears to be an excellent predictor ($r=0.986$)
    of its difficulty
    (Fig.~\ref{fig:difficulty_by_multiplicity}).
}

\subsection{Discussion}

\subsubsection{Behavior of the integer \added[id=R1]
{linear}
program}

As seen in Fig.~\ref{fig:count_by_multiplicity},
only a limited subset of our instances (342 of 10,986)
have been submitted to CPLEX.
This was for practical reasons:
despite a powerful testing environment
(Linux Ubuntu 12.04
on Intel Core i5-3570K
with 4 cores
of 3.4 GHz and 4 GB of RAM),
CPLEX turned out to need
a generous time-limit of one hour
to be able to solve to optimality
a mere 12.6~\% of this subset
(i.e., 43 instances).
The ratio dropped to 3.8~\%
when the average multiplicity reached 13;
above 20, no more success was recorded.
Thus,
this ILP quickly becomes irrelevant as the multiplicity increases,
i.e.,
as \textsc{Pagination} starts to distinguish itself from \textsc{Bin Packing}.

\replaced[id=R2]
    {Unless}
    {Until}
we could find strong valid inequalities to improve it,
our experimentations suggest
that the heuristic approach
constitutes a better alternative.

\subsubsection{Comparison of the heuristic methods}

\begin{figure}[htbp]
\begin{center}
    \includegraphics[width=\linewidth]{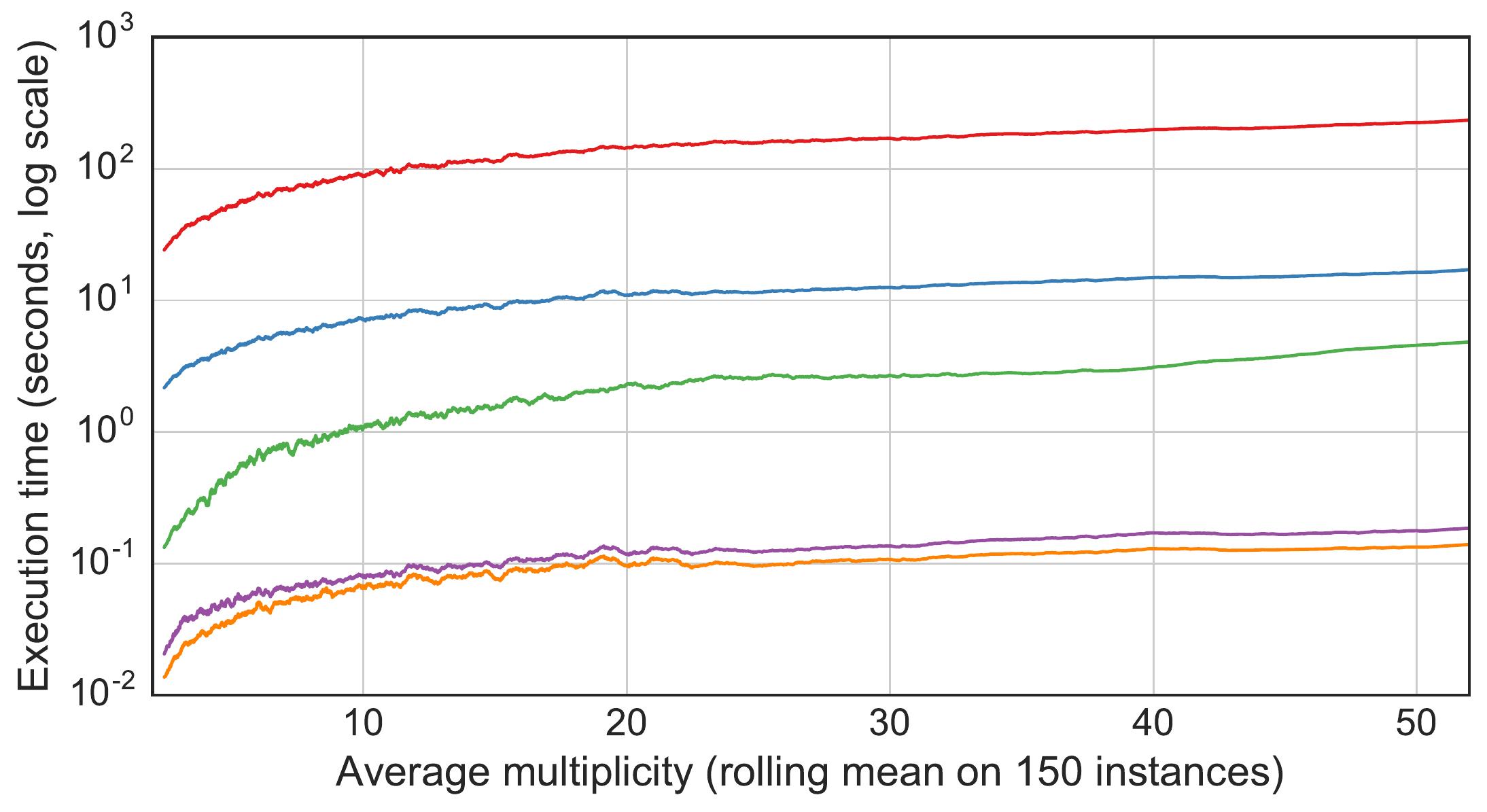}
    \caption{\label{fig:speed_by_multiplicity}\it
    Performance of the main heuristics (log scale).
    From top (slowest) to bottom (fastest):
    \textsc{Grouping GA},
    \textsc{Standard GA},
    \textsc{Overload-and-Remove},
    \textsc{Best Fusion},
    \textsc{First Fit}.
    }
\end{center}
\end{figure}

All of our heuristics have been implemented in Python 2.7,
and tested under Mac OS X 10.10
on Intel Core i5-3667U
with 2 cores\footnote{Since Python can only execute on a single core,
we usually launched two processes in parallel.}
of 1.7 GHz and 4 GB RAM.
The average execution time ranges from
less than 0.1 seconds for the greedy algorithms,
1 second for \textsc{Overload-and-Remove},
7 seconds for \textsc{Standard GA},
through
90 seconds for \textsc{Grouping GA}.
For the two GAs,
the following parameters were selected
as offering the best ratio quality/time:
80 individuals,
50 generations,
crossover rate of 0.90,
mutation rate of 0.01.
Their initial population was constituted
of valid paginations obtained
by applying \textsc{First Fit}
to random permutations of the tile set.
Figure~\ref{fig:speed_by_multiplicity} shows
how the various algorithms scale as multiplicity grows:
performance-wise at least,
all remain practical on our most difficult instances.

\added[id=Authors]
    {Note that this analysis, and the next one,
    are carried out on a moving window
    of equally-sized subsets of instances
    sorted by increasing average multiplicity:
    by eliminating noise,
    this technique helps to visually separate the different curves;
    it has the side-effect of making some endpoints disappear,
    but,
    as explained in Section~\ref{sec:avg-multiplicity},
    their loss
    is greatly outweighed
    by the gain
    of a constant confidence level
    on the remaining data.}

\begin{figure}[htbp]
\begin{center}
    \includegraphics[width=\linewidth]{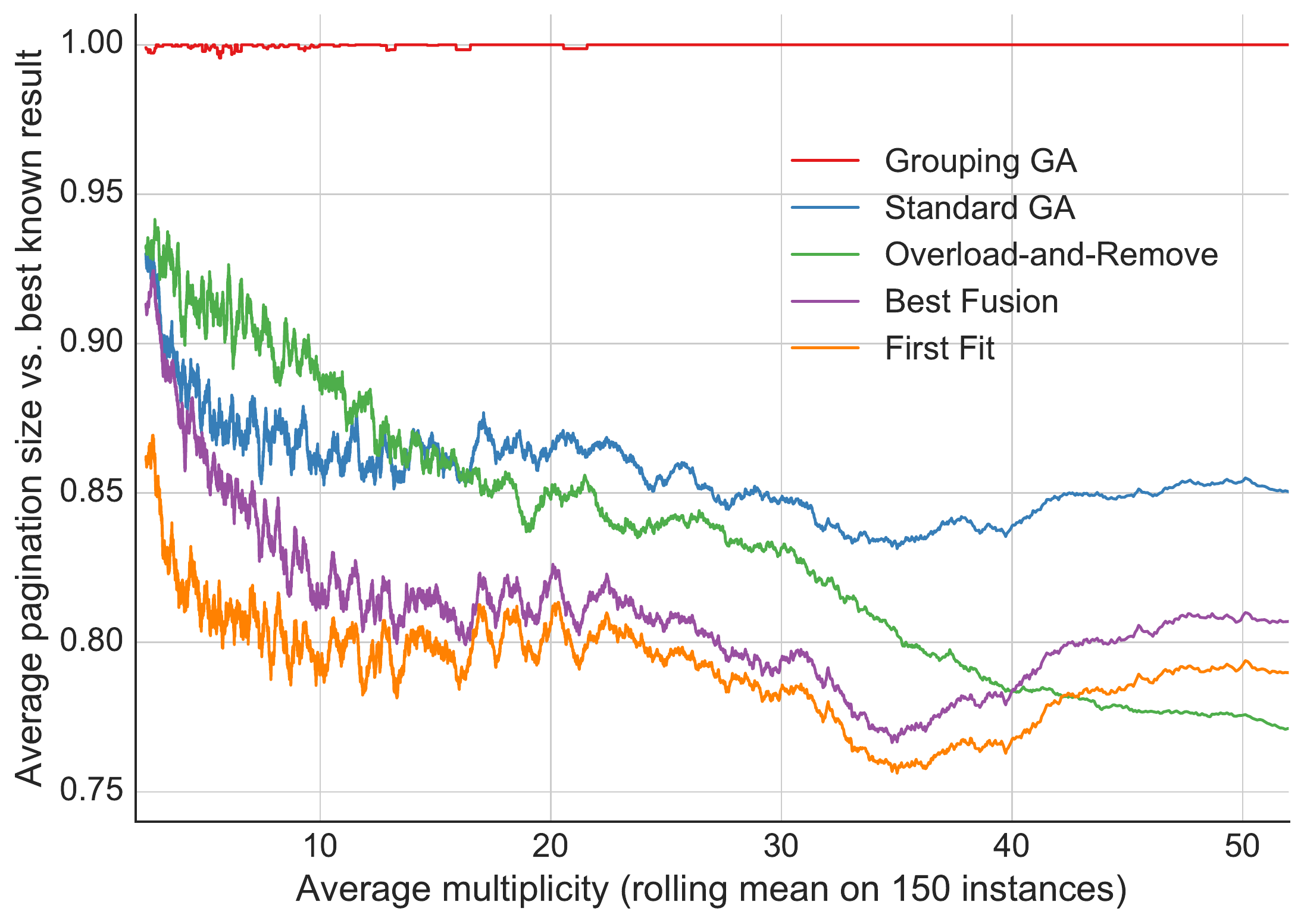}
    \caption{\label{fig:relative_size_by_multiplicity}\it
    Relative quality of the five main heuristics.
    The outcomes are plotted at $y = \frac{\mathrm{best~size}}{\mathrm{size}}$,
    with $y=1$ corresponding
    to the best \emph{known} solution
    (which is either
    the optimal or the best feasible solution found by CPLEX,
    or the smallest approximation calculated for the given instance).
    }
\end{center}
\end{figure}

Figure~\ref{fig:relative_size_by_multiplicity} compares the results
of the\deleted[id=Authors]{se} various heuristics.
The irregularities of the top line also somehow give indirect insight
into the rare achievements of our ILP \added[id=R2]
    {(among the 342 instances submitted to CPLEX,
    only 6 proved to have a better solution
    than the result of the best heuristic).}
General observations and recommendations that can be derived
from the underlying data are as follows.

\paragraph{\textsc{Standard GA} can definitely be ruled out}
As expected from Section~\ref{sec:grouping_model},
\textsc{Standard GA}
was consistently surpassed by the more sensible \textsc{Grouping GA}:
no more than 4 exceptions (0.036~\%) occurred.
Rather suspiciously, 
all of them involved a \emph{minimal} instance,
i.e., subject to a 2-page pagination.
Further examination revealed that,
in each case,
this optimal pagination was already present
in the initial random population constructed by \textsc{First Fit};
hence, its apparition was by no means
indicative of some rare small-scale superiority of \textsc{Standard GA}:
due to the fact that our implementation systematically transmits
the best individual to the next generation,
it was \emph{preserved},
rather than \emph{produced} by the evolution.
This may also partially explain as pure chance
the comparatively good performances in the lowest multiplicities;
beyond that,
the curve stabilizes quickly
around an 85~\% efficiency of \textsc{Grouping GA}.
Finally,
note that,
up to an average multiplicity of 15,
\textsc{Standard GA} is even outclassed by \textsc{Overload-and-Remove},
a six times faster heuristic.

\paragraph{\textsc{Grouping GA} produces the best overall results}
When quality is the top priority,
\textsc{Grouping GA} is the way to go:
it almost always (99.64~\% of cases) guarantees
equivalent or
smaller paginations
than any other heuristic.
ILP did improved on it in 6 cases (1.75~\% of the 342 selected instances):
2 with a better feasible solution,
4 with the optimal solution.
But even if such good surprises would multiply
on the whole set of instances,
we must keep in mind that
CPLEX was given
one hour on four cores,
against about 90 seconds
to a pure Python implementation
running on a single core (twice as slow):
\textsc{Grouping GA} has yet to unleash its full potential.

\paragraph{The fastest non-genetic contender is among  \textsc{Best Fusion} and \textsc{Overload-and-Remove}}
If speed matters,
the choice depends on the average multiplicity of the instance:
in most cases, \textsc{Overload-and-Remove} records quite honorable results.
It is even the only non-genetic algorithm which proved occasionally
(0.2~\% of cases)
able to beat \textsc{Grouping GA}.
However,
its quality regularly deteriorates
(count up to 5~\% for a 10 points increase in multiplicity).
Around 35, somewhat surprisingly,
the greedy algorithms get more and more competitive,
with \textsc{Best Fusion} taking over at 40.
Regardless of why this happens,
a specialized heuristic
for such deeply intricate instances would certainly be needed;
in the meantime,
\textsc{Best Fusion} represents the best tradeoff when average multiplicity meets or exceeds the 40 mark.

\section{Conclusion}\label{sec:conclusion}
In this paper,
we revisited an extension of \textsc{Bin Packing}
originally devised by Sindelar et al.\ \cite{Sindelar:2011}
for the virtual-machine colocation problem.
We broadened its scope to an application-agnostic sharing model:
in \textsc{Pagination},
two items can share unitary pieces of data,
or symbols,
in any non-hierarchical fashion.
We showed that
with such overlapping items, or tiles,
the familiar tools and methods of \textsc{Bin Packing}
may produce surprising results:
for instance,
while
the family of \textsc{Any Fit} approximations
have no more guaranteed performance,
genetic algorithms still behave particularly well
in the group-oriented encoding of \cite{Falkenauer:1992}.
We tested all these algorithms
on a large set of random instances,
along with an ILP \added[id=R2]{solver},
and some specialized greedy and non-greedy heuristics.
The choice of the best one is not clear-cut,
but depends on both time/quality requirements
and the average multiplicity of the symbols.
The latter measure was proposed
as a predictor of the \added[id=R2]{statistical} difficulty of a given instance,
and correlated experimentally with the actual outcome
of our various algorithms.

Obviously,
this work did not aim to close the problem,
but rather to open it to further research,
by providing
the required vocabulary,
several theoretical tools,
and an extensive benchmark.
Indeed,
numerous directions should be investigated:
examples of these are
worst-case analysis,
proof of lower bounds,
elaboration of efficient cuts,
etc.
To make \textsc{Pagination} more manageable,
a promising approach restricts
it to one single page \cite{Grange:2015, Rampersaud:2014}.

	\bibliographystyle{plain}
	\bibliography{pagination}
\end{document}